\documentclass[sigconf]{acmart}

\usepackage{booktabs}
\usepackage{multirow}
\usepackage{graphicx}

\definecolor{orang}{rgb}{0.81,0.43,0.24}
\definecolor{darkcyan}{rgb}{0.07,0.51,0.77}

\newcommand{\name}[1]{VRSight}

\newcommand{\changed}[1]{\textcolor{black}{#1}}
\newcommand{\blue}[1]{\textcolor{blue}{#1}}
\AtBeginDocument{%
  }

\copyrightyear{2025}
\acmYear{2025}
\setcopyright{cc}
\setcctype{by}
\acmConference[UIST '25]{The 38th Annual ACM Symposium on User Interface Software and Technology}{September 28-October 1, 2025}{Busan, Republic of Korea}
\acmBooktitle{The 38th Annual ACM Symposium on User Interface Software and Technology (UIST '25), September 28-October 1, 2025, Busan, Republic of Korea}\acmDOI{10.1145/3746059.3747641}
\acmISBN{979-8-4007-2037-6/2025/09}

\begin{document}

\title{VRSight: An AI-Driven Scene Description System to Improve Virtual Reality Accessibility for Blind People}

\settopmatter{authorsperrow=4}
\author{Daniel Killough}
\email{dkillough@wisc.edu}
\orcid{0009-0002-2623-0528}
\affiliation{%
  \institution{University of Wisconsin-Madison}
  \streetaddress{1210 W Dayton St}
  \city{Madison}
  \state{Wisconsin}
  \country{USA}
  \postcode{53715}
}

\author{Justin Feng}
\email{jfeng233@wisc.edu}
\authornote{Authors 2-5 contributed equally to this research.}
\orcid{0009-0003-9685-8238}
\affiliation{%
  \institution{University of Wisconsin-Madison}
  \streetaddress{1210 W Dayton St}
  \city{Madison}
  \state{Wisconsin}
  \country{USA}
  \postcode{53715}
}

\author{Zheng Xue ``ZX'' Ching}
\authornotemark[1]
\email{zzching@wisc.edu}
\affiliation{%
  \institution{University of Wisconsin-Madison}
  \streetaddress{1210 W Dayton St}
  \city{Madison}
  \state{Wisconsin}
  \country{USA}
  \postcode{53715}
}

\author{Daniel Wang}
\authornotemark[1]
\email{dwang448@wisc.edu}
\affiliation{%
  \institution{University of Wisconsin-Madison}
  \streetaddress{1210 W Dayton St}
  \city{Madison}
  \state{Wisconsin}
  \country{USA}
  \postcode{53715}
}

\author{Rithvik Dyava}
\authornotemark[1]
\email{rdyava05@gmail.com}
\orcid{0009-0006-1261-751X}
\affiliation{%
    \institution{University of Wisconsin-Madison}
    \city{Cypress}
    \state{Texas}
    \country{USA}
    \postcode{77433}
}

\author{Yapeng Tian}
\email{yapeng.tian@utdallas.edu}
\affiliation{%
  \institution{The University of Texas at Dallas}
  \streetaddress{}
  \city{Richardson}
  \state{Texas}
  \country{USA}
  \postcode{75080}
}

\author{Yuhang Zhao}
\email{yuhang.zhao@cs.wisc.edu}
\affiliation{%
  \institution{University of Wisconsin-Madison}
  \streetaddress{1210 W Dayton St}
  \city{Madison}
  \state{Wisconsin}
  \country{USA}
  \postcode{53715}
}

\renewcommand{\shortauthors}{Killough, et al.}

\begin{abstract}

Virtual Reality (VR) is inaccessible to blind people. While research has investigated many techniques to enhance VR accessibility, they require additional developer effort to integrate. As such, most mainstream VR apps remain inaccessible as the industry de-prioritizes accessibility. We present \textit{VRSight}, an end-to-end system that recognizes VR scenes \textit{post hoc} through a set of AI models (e.g., object detection, depth estimation, LLM-based atmosphere interpretation) and generates tone-based, spatial audio feedback, empowering blind users to interact in VR without developer intervention. To enable virtual element detection, we further contribute \textit{DISCOVR}, a VR dataset consisting of 30 virtual object classes from 17 social VR apps, substituting real-world datasets that remain not applicable to VR contexts. Nine participants used VRSight to explore an off-the-shelf VR app (Rec Room), demonstrating its effectiveness in facilitating social tasks like avatar awareness and available seat identification.

\end{abstract}




\begin{teaserfigure}
    \centering
    \includegraphics[width=\textwidth]{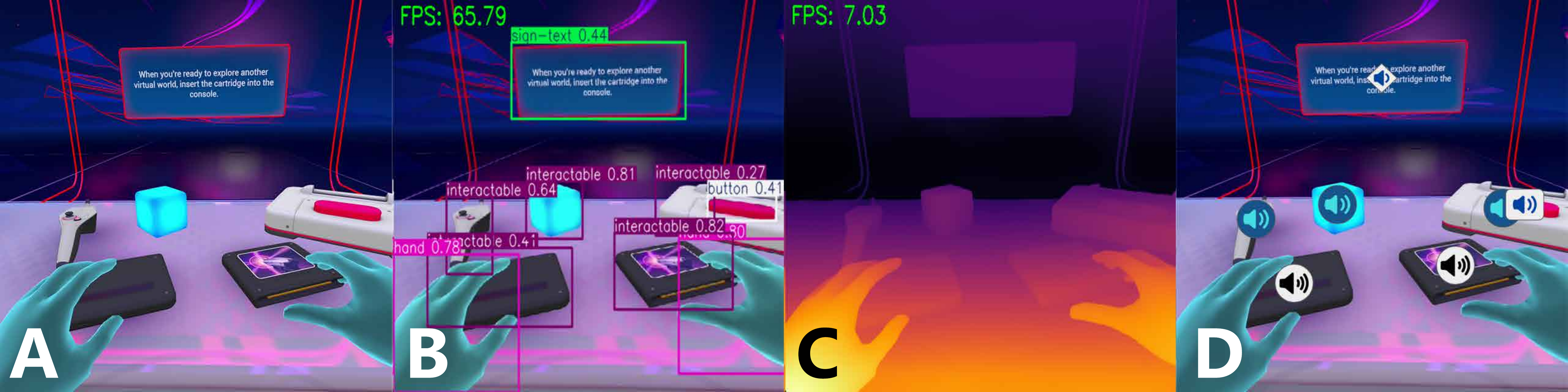}
    \vspace{-5ex}
    \caption{VRSight is a \textit{post hoc} 3D screen reader for blind VR users: (A) the original VR scene, (B) the fine-tuned object detection, (C) depth map of the VR scene, and (D) spatial audio representation based on AI recognition of the scene. }
  \Description{This image shows a sequence of four VR frames. The first frame is a VR scene that displays a pair of virtual hands in the foreground, with some objects like a game controller, a blue cube, and a console on a table. A text box in the background provides instructions about the game, reading: ``When you're ready to explore another virtual world, insert the cartridge into the console''. The second frames shows the same scene with object detection overlays. A few bounding boxes highlighted the identified objects in the scene, labeled as ``sign-text'', ``hand'', and ``interactable'' respectively, with corresponding confidence scores. The third frame shows a depth map of the scene, where hands and objects are shaded with colors representing their distance from the viewer. The fourth frame demonstrates the text-to-speech feature. A few sound icons are placed above the objects, indicating interactive audio feedback and text-to-speech functions, with objects closer to the user showing as louder.}
    \label{fig:teaser}
\end{teaserfigure}

\maketitle

\section{INTRODUCTION}

Virtual Reality (VR) platforms enable diverse immersive experiences, and rapidly increasing hardware capabilities and computational power are causing VR to evolve from a niche technology to a mainstream tool in areas like gaming~\cite{dani2019impact}, socializing~\cite{maloney2020falling,zhang2023diary}, education~\cite{hu2017virtual,kavanagh2017systematic}, and
healthcare~\cite{mantovani2003virtual}.
Despite major advances in VR hardware and software, VR accessibility (a11y) for blind and low vision people (BLV) has not kept pace. VR devices like the Meta Quest and Apple Vision Pro currently rely almost exclusively on visual prompts such as on-screen text, visual gestures, and color-coded objects to denote interactivity, all of which create significant barriers for BLV~\cite{teofilo2018evaluating,trewin2009exploring,wong2018vr}.

Standardized visual a11y features for VR do not yet exist besides basic adjustments like color filters and text size~\cite{metaquestaccessibility}. A11y features for users needing support outside of colorblindness or magnification have yet to be fully introduced (e.g., for blind users, people with central vision loss, people with blurred vision); furthermore, VR developers see people with visual impairments as unable to use virtual reality devices at all, and are therefore largely resistant to supporting them~\cite{killough2024xr}. Some a11y research for 3D virtual environments has progressed for common BLV difficulties like in-scene navigation~\cite{nair2021navstick} and movement~\cite{ribeiro2024investigating}, but research lacks an end-to-end description system to describe virtual objects, like a 3D screen reader. Furthermore, existing VR research largely requires developer intervention in a controlled environment, but users need a post hoc solution that works in already-released applications without additional developer effort~\cite{killough2024xr}. 

Therefore, we present VRSight, an end-to-end automated object detection system that leverages state-of-the-art machine learning models to interpret and translate real-time spatial information in virtual worlds through 3D audio. We also present DISCOVR, a dataset sourced from 17 free, top-selling VR apps in over 30 categories of VR objects (\textit{e.g.,} interactables, avatars, seats, signs). VRSight runs post hoc on the VR headset's visuals to automatically identify virtual objects based on these object classes; process the detected visuals to generate audio effects; combine the data with a depth map to spatialize the generated audio; and relay information back to users. \changed{This post hoc method allows VRSight to run as an overlay on existing apps, capturing and analyzing visual frames without requiring specific VR devices nor developer integration.} VRSight also detects the scene's tone and customizes the audio descriptions accordingly to enhance user immersion.

We evaluated VRSight with a thorough technical evaluation of DISCOVR, a generalized evaluation citing face validity, and a user study with nine BLV participants. VRSight runs on standalone and PC VR headsets with an end-to-end latency as little as \char`~2.1ms from keypress to audio playback. BLV participants evaluated VRSight as mostly accurate, helpful, and easy to use \changed{in exploration-based social VR contexts. They gave insightful feedback towards the creation of future interaction methods, and we invite future work to build on top of VRSight's default user interaction methods, developing additional effective, user-informed audio description techniques.}

VRSight contributes the following:
\begin{itemize}
    \item An end-to-end system combining multiple AI models to synthesize tone-relevant spatial audio descriptions from the headset's visuals. Our system \changed{runs} post hoc on any VR application without requiring app developer intervention. 
    \item The DISCOVR dataset including 17,691 labeled VR images focused on 17 free, top selling VR apps and a corresponding fine-tuned, state-of-the-art object detection model trained on the custom dataset, supporting real-time virtual object detection for 30 VR object categories. 
    \item Evaluations of VRSight's performance in isolation and in nine interviews with end-users, including suggestions for future interaction methods using VRSight.
\end{itemize}

\section{RELATED WORK}
\subsection{VR Challenges for BLV}
VR a11y for BLV has been widely studied, with significant challenges as VR environments largely rely on visual feedback. Despite major advances in VR hardware and software, standardized visual a11y features for VR remain mostly limited to basic adjustments like color filters and text size~\cite{metaquestaccessibility}. 
Prior work has explored approaches to make VR more accessible to BLV, primarily through audio-based solutions including techniques for exploring 2D objects~\cite{heuten2006city} and interfaces~\cite{oliveira2015music}, conveying 3D virtual environments~\cite{balasubramanian2023enable,lokki2005nav, nair2021navstick, nair2022uncovering}, supporting avatar awareness~\cite{ji2022vrbubble}, and navigating simulated real-world scenarios~\cite{walker2006navigation}. 

Navigation and spatial understanding represent particular challenges for BLV in VR. Nair et al. \cite{nair2021navstick, nair2022uncovering} addressed these challenges by repurposing the right VR controller's thumbstick from camera control to a spatial audio reader in an indicated direction. 
In a similar vein, Waters and Abulala \cite{Waters2001bat} simulated echolocation to help BLV independently explore virtual environments. 


Communicating social contexts in VR also presents a significant challenge. To address this issue, Ji et al.~\cite{ji2022vrbubble} designed an audio-based interaction technique to notify users of surrounding avatars based on social distance. 
Beyond audio feedback, researchers have also utilized haptics to enable BLV to explore and navigate virtual spaces~\cite{schloerb2010blindaid}, or interact with VR objects~\cite{jansson1999haptic, tzovaras2009interactive, tzovaras2002design, wedoff2019virtual}.
For example, Jung et al.~\cite{jung2024accessible} combined a series of haptic and audio cues to notify PVI of nonverbal social cues like eye contact, head nodding, and head shaking. Zhao et al.~\cite{zhao2018canetroller} created a wearable VR controller based on a brake mechanism to simulate white cane interaction for PVI in VR, which Siu et al.~\cite{siu2020cane} further improved with audio feedback to emulate echolocation. Such approaches demonstrate how multimodal feedback may enhance environmental understanding for BLV in virtual spaces. 

While controller input systems~\cite{nair2021navstick} and locomotion methods~\cite{ribeiro2024investigating} have been explored for VR navigation, complex scene understanding remains underdeveloped. Some research has begun exploring 3D screen reader-adjacent approaches, like Herskovitz et al. projecting 3D images onto 2D screens for VoiceOver~\cite{herskovitz2020making}, but these may have platform limitations.
Dang et al.~\cite{dang2023opportunities} proposed an approach that ``focuses on conveying visual information in a virtual space through multimodal feedback such as audio description/feedback, haptic feedback, and spatial audio'', but comprehensive implementation of such systems in VR remains limited. A significant gap in research exists for end-to-end description systems that screen read 3D virtual environments.



\subsection{Audio A11y Techniques for BLV}
Audio-based accessibility techniques for blind and low vision (BLV) users have been widely studied, particularly outside of VR. Audio descriptions (AD) have been implemented across a wide range of formats, from traditional pre-recorded videos (e.g., films~\cite{snyder2005audio,incredibles,branje2012livedescribe,adpguidelines}, YouTube videos~\cite{adp}), to more recent formats like 360\textdegree~ video~\cite{chang2022omniscribe,hara2015improving,jiang2023beyond}, livestreams~\cite{jain2023front,killough2023exploring}, and short-form video~\cite{van2024making}.





Many systems rely on manual AD authoring through an interface~\cite{youdescribe, killough2023exploring} or connecting with a sighted volunteer~\cite{bemyeyes}, but recent work has explored automated approaches to generate~\cite{jain2023front} and edit~\cite{pavel2020rescribe} audio descriptions. \changed{Matching the text-to-speech (TTS) tone to narrative or emotional context has also been shown to reduce distraction and enhance \textit{immersion} in both film~\cite{bittner2012audio} and video~\cite{pavel2020rescribe}: a key tenet of VR.}

\subsubsection{\changed{Spatial Audio for BLV Navigation and Environmental Awareness}}
\changed{Traditional 2D screen readers like NVDA and JAWS fail to convey object location or spatial relationships critical for immersive navigation in 3D environments. To address this problem, prior work like Chang et al.'s OmniScribe~\cite{chang2022omniscribe} used directional audio to describe objects in 360\textdegree~ videos, providing directional descriptions to enhance immersion.} Smith et al.~\cite{smith2018rad} developed the Racing Auditory Display (RAD) system with a ``sound slider'' for trajectory information and a ``turn indicator system'' for upcoming turns, demonstrating how spatial audio cues can convey critical, game-specific information. Research has also explored spatial audio for virtual conferencing environments~\cite{hyrkas2023spatialized,nowak2023hear} \changed{and virtual navigation~\cite{afonso2022spatial,clemenson2021rethinking}, with users developing accurate cognitive maps from these descriptions. Additional consistent spatial metaphors like left-to-right object ordering have been shown to improve orientation~\cite{coughlan2020icchp,sato2025audio}.} 





\subsubsection{\changed{Audio Modalities and Levels of Detail}}
\changed{Prior work also offers guidance on optimizing audio feedback content. The ``Less is More'' principle has been validated across multiple studies: concise, structured, and user-controllable descriptions are generally preferred over verbose narration~\cite{fresno2014less,natalie2024audio}, but this principle applies differently across contexts. 
Sound effects and spatial audio excel for navigation and environmental awareness, while verbal descriptions remain essential for semantic information and object identification. VR research demonstrates that well-designed spatial audio can provide equivalent spatial information to real-world acoustic environments.}


\changed{A meta-analysis of auditory interface modalities~\cite{nees2023auditory} ranks speech and spearcons highest for recognition accuracy and speed, followed by auditory icons and earcons. In spatial navigation tasks, however, spatial audio consistently outperforms verbal-only cues~\cite{afonso2022spatial,clemenson2021rethinking}, leading to faster task completion and improved spatial awareness. These insights inform our hybrid use of speech, spatial panning, and adaptive verbosity in VRSight's audio descriptions.}

\subsection{Proxy-Based and Post Hoc A11y Tools}
\changed{Prior work shows VR developers often de-prioritize a11y (e.g., due to time/effort constraints~\cite{killough2024xr}).}
To address gaps in existing technologies without a fundamental app redesign, researchers in web contexts have developed post hoc modifications using overlays or proxy layers \changed{as crucial fallbacks when developer integration is infeasible}. Such an approach has been widely applied to web contexts~\cite{bigham2008webanywhere,takagi2000transcoding}, where proxies improve accessibility by adding enhanced functionality through intermediate layers. Similar approaches have been used in mobile contexts, with interaction proxies modifying the input and output of underlying applications to enhance a11y~\cite{zhang2017interaction}. \changed{However, post hoc a11y modifications for VR remain largely unexplored; Zhao et al.'s SeeingVR~\cite{zhao2019seeingvr} may be the only system that incorporates post hoc methods.} SeeingVR includes several post hoc modifications that could make existing VR applications more accessible without requiring developers to rebuild their apps from scratch; however, SeeingVR primarily targets low vision rather than blind users, who may require more comprehensive scene understanding and object recognition. Furthermore, despite the \textit{promise} of proxy-based solutions, few comprehensive systems exist for VR a11y that function without developer intervention in existing applications. Current VR research largely requires developer implementation in controlled environments, but as Killough et al.~\cite{killough2024xr} note, users need post hoc solutions that work in already-released applications without additional developer effort.


\changed{VRSight directly addresses this need by functioning as a user-side, post hoc overlay system aligning with the call for solutions that can be applied to already-released applications, making mainstream VR experiences more accessible for BLV.} To our knowledge, VRSight is one of the first proxy systems for VR to comprehensively combine real-time object detection, spatial audio, and scene-aware tonal descriptions to enhance accessibility for BLV across different VR platforms and applications.

\section{SYSTEM IMPLEMENTATION}

To facilitate VR scene understanding and exploration for BLV, we designed and built VRSight, a post hoc software tool that captures visual frames from a VR application in real-time and generates spatialized sound effects and speech based on the detected VR objects at simulated 3D locations. 
Below we discuss each component of the system, including (1) the collection and annotation process for the VR object dataset; (2) fine-tuning an object detection model on our VR dataset; and (3) building a post hoc system with a set of default interaction methods that we evaluated with BLV. \changed{The labeled dataset \textit{(DISCOVR)}, fine-tuned model weights \textit{(DISCOVR-best.pt)}, and post hoc system \textit{(VRSight)} are open-sourced at \blue{\href{https://github.com/MadisonAbilityLab/VRSight}{https://github.com/MadisonAbilityLab/VRSight}}. Our current implementation uses the Meta Quest 3 VR headset for popularity and cost accessibility, but other platforms with mirroring utilities (\textit{e.g.,} Valve Index/SteamVR) can be supported.}





\subsection{Dataset and Detection Model Development} 
\label{sec:dataset}
Powering VRSight, we present DISCOVR---\textit{DIgital Social Context Objects in VR}---a dataset of labeled objects in social VR scenes.
Research has investigated using computer vision tools like object detection to describe real-world environments~\cite{venkata2020research}, but existing models trained on real-world datasets (\textit{e.g.,} COCO~\cite{lin2014microsoft}) inconsistently or seldom identified digital objects in VR contexts. This is likely due to the domain gap between real and virtual environments---VR objects often differ in texture, lighting, rendering style, and context, even when they mimic real-world counterparts like televisions or benches.
To address this gap, we need to create our own dataset tailored for VR contexts. 

\subsubsection{Data Collection.} To collect images for DISCOVR, we grounded our efforts on the most popular standalone consumer VR headset line (Meta Quest) and VR apps, focusing on social VR for its wide diversity in worlds, types of objects, and low barrier of entry. We conducted an exhaustive search on the Meta Quest store~\cite{meta_experiences} for free, ``top selling'' applications for socializing with other users in virtual reality. 
Our search returned 24 free social VR apps, which we further narrowed to omit gambling (\textit{e.g.,} PokerStars), apps that are strictly games (\textit{e.g.,} Gorilla Tag), 2D apps (\textit{e.g.,} Facebook), apps with primarily real-life visuals (\textit{e.g.,} Wooorld), and those requiring subscriptions after a free trial period (\textit{e.g.,} Villa, Immerse, Innerworld). 
Our trimmed list included 15 social VR apps, including VRChat, Rec Room, ROBLOX, Remio, Half + Half, Flipside, Alcove, Engage, Spatial, Meta Horizon Worlds, Zoe, vTime XR, MeetinVR, Multiverse, and Arthur.
We decided to additionally include two official platform demos, Oculus First Steps and Oculus First Contact, to provide a standard for the types of objects common in VR (\textit{e.g.,} interactables, signs, non-human avatars) and support these apps that automatically open on setup for older VR devices like the Oculus Rift and Quest 1. \changed{Including these official demos resulted in a final total of 17 VR apps.}

Six researchers recorded VR screen recordings using Quest 2, Quest Pro, and Quest 3 and uploaded still frames to Roboflow~\cite{roboflow_homepage}, a popular online tool for annotating and training computer vision models. \changed{Each researcher actively played several social VR games, recording at least 20 minutes across at least three different environments per application (excluding Oculus First Contact, which is a single scene). To minimize annotating similar frames in the dataset, we extracted one frame per second from the screen recordings and randomized the upload order, resulting in a total upload count of approx. 25,000 images across the 17 VR apps.} 

\subsubsection{Data Annotation \& Augmentation.}
Researchers then independently used Roboflow to manually annotate a starting set of 300 images each, or 60 images from Oculus First Steps, Oculus First Contact, VRChat, Rec Room, and ROBLOX, annotating any VR object they felt was necessary using open coding, adding new labels to a list. Researchers annotated their own VR screen recordings to reduce the chance for ambiguity in potentially unfamiliar apps. After completing 300 images each, researchers met to compare lists and agreed upon an initial master list of 40 classes. 
Four researchers then annotated approximately 2000 images each from their recordings, with another researcher reviewing annotations to reduce errors and ensure consistency within classes. During the annotation process, and after training and testing an initial model, researchers met to refine the master class list, combining classes together (\textit{e.g.,} \textit{``spawner-players''} and \textit{``spawner-items''} combined to \textit{``spawner''}) and removing classes seldom identified by the model despite sufficient training examples (\textit{e.g.,} \textit{``pointer-target''}, a colored circle indicating the end of a user's VR laser pointer). Our final class list consists of 30 types of VR objects, enumerated in Table~\ref{tab:classes}. 

\begin{figure}
    \centering
    \includegraphics[width=1\linewidth]{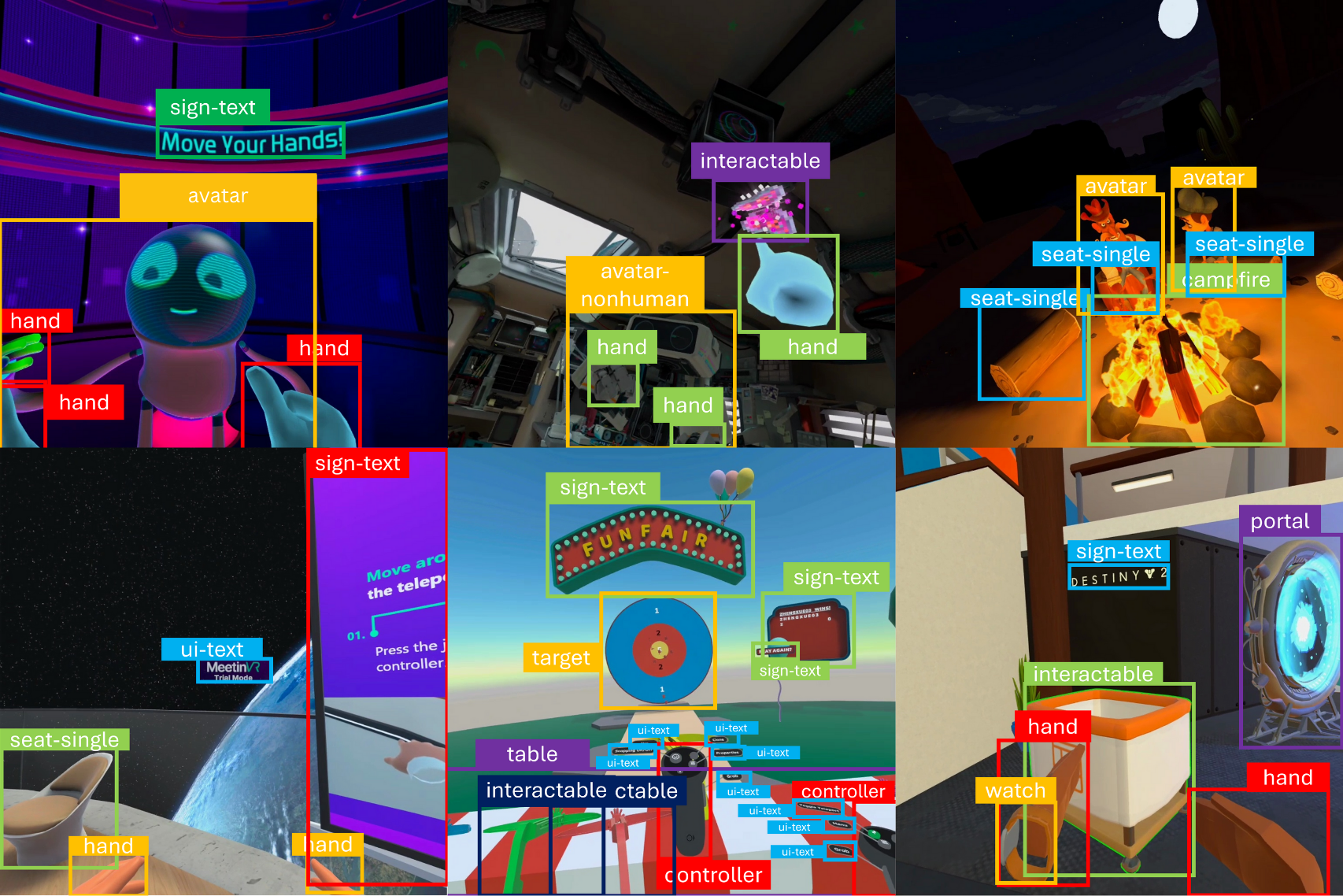}
    \caption{Illustrated example annotations from the dataset.}
    \label{fig:datasetexamples}
    \Description{A grid of six virtual reality interface screenshots showing labeled interactive elements. The images display various VR environments with computer vision object detection highlighting components such as: avatars (human and non-human), hands, controllers, interactive objects, seats, tables, portals, targets, and text signs. Each element is color-coded with bounding boxes (red, green, blue, orange, and purple) identifying interface components that users can interact with in the virtual environment.}
\end{figure}

After annotating, we split these into a standard 70-20-10 split: 70\% of the data for training, 20\% for validation, and 10\% for testing, ensuring that duplicate images did not appear in multiple splits. 
To help verify the generalizability of our model, we randomly excluded three apps from training (Engage, Spatial, and Half+Half), including them only in validation and testing with a 67-33 split at the same absolute image quantity as other apps. \label{split-training}
We then used Roboflow's generate augmentation tool to create 3 image augmentations per image in the training split, randomly combining a series of horizontal or vertical flips, rotation 0--360\textdegree, 0--20\% magnification, and ±15\textdegree~ horizontal and/or ±15\textdegree~ vertical shear to vary our training data and help account for differences in how people use their headset~\cite{roboflow_augmentation}. We also resized images to 640x640 per the recommendations of our object detection model. This process resulted in a total of 17,691 annotated images across all 17 VR applications and a new distribution of 15,207 training images, 1,645 validation images, and 839 test images. We illustrate some example annotations from our dataset in Figure~\ref{fig:datasetexamples}. 

\subsubsection{Model Fine-Tuning.}
To use the DISCOVR dataset with VRSight, we fine-tuned a YOLOv8n~\cite{yolov8_ultralytics} object detection model on DISCOVR using an NVIDIA A100 GPU for 250 epochs. YOLOv8 is a state-of-the-art model offering a good balance of high accuracy and speed for our use case. 
We opted to fine-tune rather than train from scratch to leverage our dataset while taking advantage of YOLOv8's proven pre-trained weights and backbone. This process enabled us to adapt the model to recognize our VR-specific classes while maintaining the robust feature extraction capabilities of the original model. 
We evaluate the performance of the model in \S~\ref{model-performance}.


\subsection{VRSight's System Flow} 
To implement VRSight, we aimed to maximize three primary components: (1) the awareness of VR objects (\textit{e.g.,} with our object detection model); (2) effective facilitation of user interaction (\textit{e.g.,} edge detection); and (3) the sense of immersion (\textit{e.g.,} spatial audio, scene-relevant speech tones).

\subsubsection{\changed{Design Justification}}
\changed{Most VR devices use visuals, audio, and haptics~\cite{bamodu2013virtual} with many applications primarily relying on visuals to convey information. To enable more accessible immersive experiences for blind and low vision (BLV) users, we designed VRSight to attempt to replace vision with audio as the primary modality. 
Haptic feedback remains infeasible for widespread deployment without specialized external hardware, since standalone platforms like Meta Quest still restrict multiple applications from running simultaneously on the headset, which limits our ability to layer haptics on the included controllers post hoc. 
Only recently do standalone devices like the Meta Quest enable background audio playback support natively on-device\footnote{\url{https://www.meta.com/blog/meta-quest-v66-software-update-reduced-passthrough-distortion-background-audio/}}. We therefore prioritized an audio-based approach that users could deploy with current hardware.}

\changed{We describe VRSight as a ``3D screen reader'' (or ``3D scene reader'') to distinguish it from 
traditional ``2D screen readers'' (\textit{\textit{e.g.,}} NVDA~\cite{nvda}, JAWS~\cite{jaws}) which provide linear audio descriptions for web content. Conversely, ``3D screen readers'' must convey environments where objects are scattered throughout space. In such environments BLV users need to know not only what objects exist, but where--- left, right, and how far away--- to navigate and interact effectively.} 
 \changed{VRSight conveys these spatial relationships using stereo panning for horizontal direction and volume scaling for distance, building on prior work demonstrating spatial audio as improving navigation and task performance for BLV users~\cite{ribeiro2024investigating}. We sort objects left-to-right to minimize confusing audio jumps, supported by Sato et al.'s recent research on localization from spatial relative sound order~\cite{sato2025audio}.
By integrating these techniques, we aim to provide intuitive, immersion-preserving audio cues without requiring external hardware or developer implementation.}

\subsubsection{System Components}
\label{syscomponents}
To achieve each goal in a post hoc system, VRSight integrates multiple AI models which each serve a specific purpose in translating visual information to audio feedback:
(1) \textbf{A Fine-Tuned Object Detection model} \textit{(YOLOv8~\cite{yolov8_ultralytics})} ~identifies 30 classes of VR-specific objects (\textit{e.g.,} avatars, buttons, signs) with bounding boxes and confidence scores. This model provides the foundation for all subsequent a11y functions.
(2) \textbf{A Zero-Shot Depth Estimation Model} \textit{(DepthAnythingV2~\cite{depth_anything_v2})} ~generates depth maps of our VR scene to understand the spatial arrangement of objects. Using this map in conjunction with output from our object detection model helps accurately position spatial audio in 3D space, communicating both what objects exist and where each exists in 3D space relative to the user. 
(3) \textbf{A Multimodal Large-language Model} \textit{(GPT-4o)} ~for two critical functions beyond our object detection model: (A) GPT analyzes the overall context and mood to set an appropriate tone for our TTS, helping BLV understand the intended emotional context of scenes to enhance immersion \changed{(``LLM-based atmosphere interpretation'')}; 
(B) GPT provides in-depth descriptions for objects with fine detail and non-text graphics like icons or emoji. 
(4) \textbf{An Optical-Character Recognition (OCR) Model} \textit{(Microsoft Azure AI Vision)} ~to read classes with text embedded in the virtual environment. Text is ubiquitous in VR interfaces but is particularly challenging for BLV. OCR reads text-heavy elements like signs, notifications, and avatar names to ensure this critical information for navigation and interaction remains accessible. We pass all generated text to:
(5) \textbf{A Text-to-Speech Generation System} \textit{(Microsoft SpeechSynthesizer)} ~that generates speech in a selected tone (\textit{e.g.,} neutral, cheerful, sad, fearful, urgent), enabling VRSight to couple object information with emotional context to enhance immersion. 

\begin{figure*}[!ht]
    \centering
    \includegraphics[width=1\linewidth]{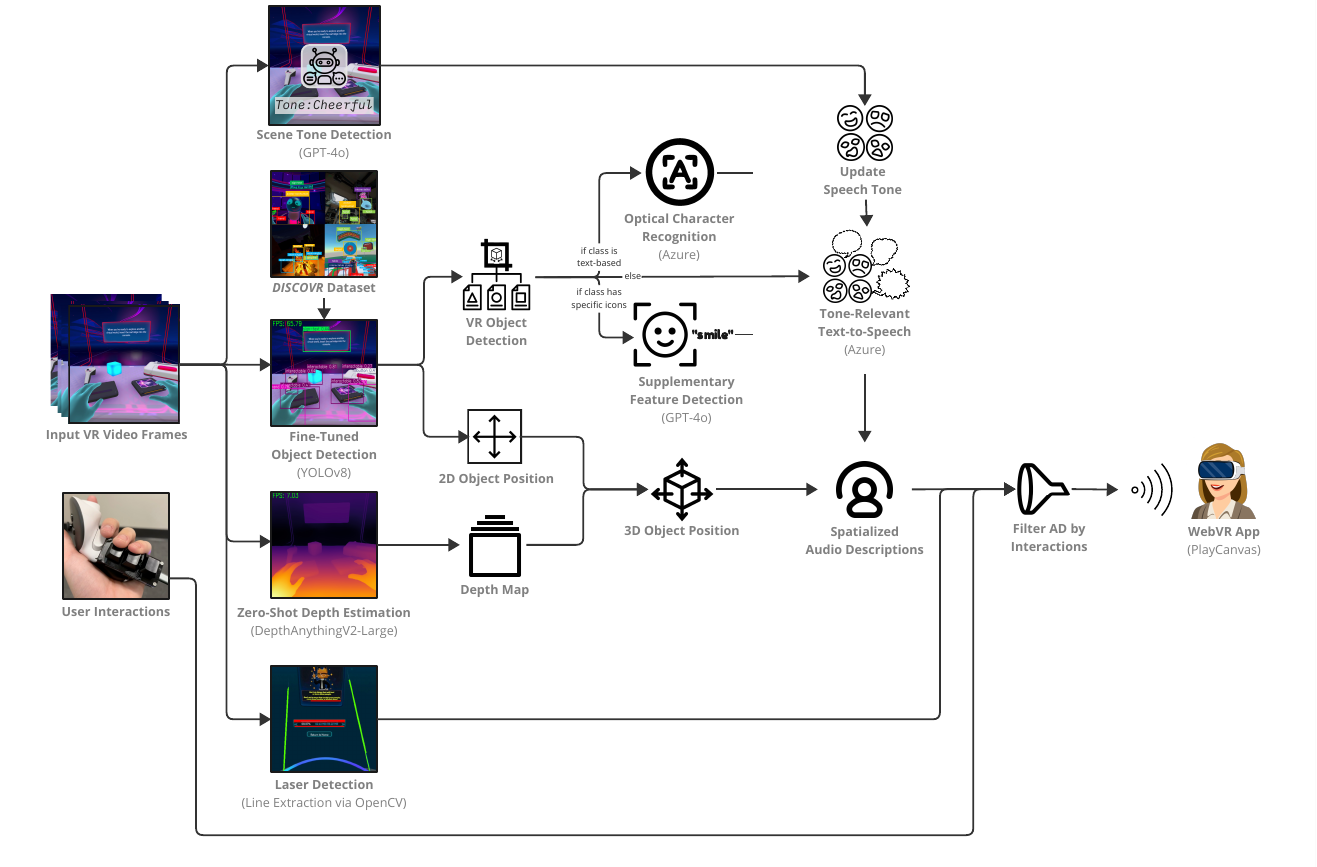}
    \caption{Simplified system flowchart showing the paths from the input VR video frames to the output webVR spatial audio.}
    \label{fig:system-flow}
    \Description{This flowchart illustrates a pipeline for processing input VR video frames. The process begins with capturing VR video frames, which are analyzed through a scene tone detection module (GPT-4o) to analyze the scene's tone (e.g., neutral). Simultaneously, the video frames are used to create a dataset for training and fine-tuning a YOLOv8 object detection model, which generates 2D bounding boxes around detected objects and classifies detected objects. A DepthAnythingV2-Large model estimates object depth to calculate their 3D positions. Depending on the object's class, further processing may occur: text-based objects undergo optical character recognition using Azure, and objects with specific icons are analyzed using a supplementary feature detection module (GPT-4o) to extract relevant features (e.g., detecting a "smile" icon as a "smile"). The extracted text or detected features are converted into speech via Azure's Text-to-Speech service and customized using the speech tone. The processed text to speech are then combined with the 3D object positions to render spatialized audio descriptions. Finally, the data is filtered via user interactions and sent to a WebVR application to provide accessible output.}
\end{figure*}

We then send all speech, sound effects, and 3D location coordinates over websocket to a WebVR utility \textit{(PlayCanvas)} to convey directional audio to the user. 
Figure~\ref{fig:system-flow} demonstrates a flowchart of how each component impacts one another.

While the object detection model identifies a series of objects in each frame, the depth model generates a map with the relative distance of each object in the virtual environment. Combining these factors enables us to accurately place objects in simulated 3D space. 
Our GPT-4o, TTS, and OCR services all contact Microsoft Azure. Although contacting a cloud service adds latency, we tested some on-device computing methods and found Azure granted higher accuracy than on-device computing methods, particularly for GPT 
and OCR. Furthermore, Azure contains many options for text-to-speech emotions to help us convey multiple scene tones to our users using the same base voice. 

\subsection{Default User Interactions}
\label{interactions}
\begin{figure*}[!ht]
    \centering
    \includegraphics[width=1\linewidth]{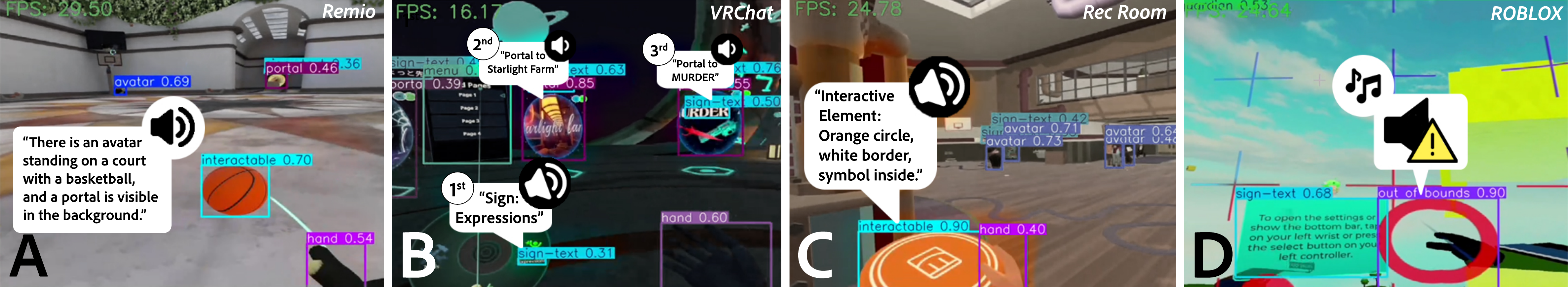}
    \caption{\changed{Demonstrating interaction methods against object detection output. (A): \textit{ContextCompass} with an overall, general scene description in Remio. (B): \textit{SceneSweep} reading out object-level descriptions from left (1st) to right (3rd) in VRChat. (C): \textit{AimAssist} providing a detailed description of the interactive element held in the user's hand in Rec Room. (D): \textit{SafeGuard} playing a spatialized warning sound when the guardian (plus symbols) or out of bounds (red circle) objects are detected.}}
    \label{fig:defaultinteractions}
    \Description{Virtual reality interface screenshots labeled A, B, C, and D showing object detection in a VR environment. Image A shows a basketball court with labeled avatar, basketball (marked as ``interactable''), hand, and portal elements, with an audio description popup. Image B shows a room with a menu and a number of labeled portals. Three objects are labeled: marked 1st is "Sign: Expressions", 2nd is "Portal to Starlight Farm", and 3rd is "Portal to MURDER." Image C shows a room with multiple avatars and an interactive element described as ``Orange circle, white border, symbol inside''. Image D displays a settings instruction panel with various labeled interface elements including sign-text and out-of-bounds markers, with audio and warning icons visible.}
\end{figure*}

VRSight works alongside the user's VR headset on a separate computer, allowing seamless integration for any VR application and device with minimal additional setup. 
While VRSight's detection system identifies virtual objects in real-time, playing all detected items for each frame may be overwhelming. Therefore, we developed four default interaction features to help evaluate VRSight with end-users, \changed{whose output is illustrated in Figure~\ref{fig:defaultinteractions}}. Each feature is activated using a three-key keyboard we attached to the left controller, ordered from most general to most specific: \\(1) For full-scene descriptions, users activate \textbf{\textit{ContextCompass}} by pressing the topmost key closest to the VR controller's grab button. \textit{ContextCompass} describes the current view as a concise, contextual summary of the scene. This query is intended to help users orient themselves in the VR scene (\textit{\textit{e.g.,}} after teleporting). \\(2) To learn more about individual objects in their environment, users activate \textbf{\textit{SceneSweep}} by pressing the middle key on the keyboard. \textit{SceneSweep} 
takes a snapshot of the current scene, then scans the entire VR environment horizontally from left to right, playing spatial audio cues corresponding to each detected object's 3D location and a classification of the type of object (\textit{e.g.,} ``sign''). \changed{\textit{SceneSweep} analyzes the user's current view and reads identified objects sorted left-to-right using spatial audio; we note that \textit{SceneSweep} does not automatically pan around the visual scene outside of the user's current field of view.} \\
(3) VR applications commonly use pointers or users' hand positions to interact with the environment. To get more detail about objects near users' hands, controllers, or laser pointers, activate \textbf{\textit{AimAssist}} with the bottom key. \textit{AimAssist} reads elements near the pointer without needing to point directly at each item. If the user triggers \textit{AimAssist} while the pointer is over a menu, it reads out precise elements present around the end of the pointer. 
Should pointers be unavailable, \textit{AimAssist} defaults to reading out objects near the user's hand or controller, including greater detail about their color or shape, or infers what the object may be.\\
(4) Finally, to aid user safety, VRSight uses \textbf{\textit{SafeGuard}}, a proactive feature designed to keep users from moving out of the VR system's defined boundary. Upon system setup, VR users may define a virtual boundary around their physical playspace--- ordinarily if users move outside their playspace the VR system shows a visual ``cage'', but \textit{SafeGuard} identifies the visual cage and issues an auditory warning, prompting users to back up. \\ 

\changed{Three interactions follow a general-to-specific approach inspired by North and Schneiderman's visualization paradigms (overview, zoom/filter, details-on-demand)~\cite{north2000snap}: \textit{ContextCompass} provides scene overviews; \textit{SceneSweep} offers scanning based on traditional screen reader navigation plus spatial audio; and \textit{AimAssist} enables precise targeting using established VR pointer paradigms. To our knowledge, \textit{SafeGuard} is a novel feature to address user safety, as current VR boundary systems are visual-only.} Each of these user interactions combines the components listed in \S~\ref{syscomponents} to provide different types of spatial awareness and information access. All features using speech (1, 2, 3) are read out in an application-relevant tone.

\section{TECHNICAL EVALUATION}
To evaluate VRSight, we first conducted a technical evaluation of VRSight's components including numbers for overall system latency and the performance of the dataset. Our findings demonstrate that VRSight is capable of accurately identifying many classes of social VR objects with real-time feedback.

\subsection{Methods}
\subsubsection{PC Requirements.} VRSight is primarily contained within a single Python program run on a connected computer. VRSight simply requires a VR headset and a connected computer with a capable-enough GPU to run the machine learning models. Since many users purchase standalone VR headsets due to their lesser overall cost and portability, we focus on standalone VR devices for these metrics (Meta Quest 3), but PCVR headsets (e.g., Pimax Crystal Light) and standalone headsets connected via Link cable are compatible provided that the PC contains a sufficiently performant GPU or a second GPU to run simultaneously without severe drops in performance. We report our benchmark numbers using an NVIDIA RTX 4090 GPU, but we have successfully run VRSight on a gaming laptop (Alienware M15) with an onboard RTX 3070 mobile GPU and achieved similar latency figures but with lower framerates. 

\subsubsection{Dataset Evaluation.} After generating 3 augmented images for each image in the training split (per the steps in \S~\ref{split-training}), our dataset, DISCOVR, contains 17,691 annotated images. 
After excluding Engage, Spatial, and Half+Half from the train split and generating 3 augmentations for the train split, the train split contained 15,207 images and 79,548 object instances. The validation split contained 1,645 images and 8,235 object instances, and the test split contained 839 images and 4,257 object instances. To evaluate the dataset further we run Ultralytics' testing API on the test split and report the average precision at 50\% Intersection over Union (IOU), 75\%, and overall (averaged between 50--95\% IOU).

Researchers met to qualitatively categorize the classes, determining six categories for the 30 VR object classes: Avatars, Informational, Interactables, Safety, Seating Areas, and VR System.

To evaluate the object detection, depth estimation, and edge detection models, we recorded 1 minute of each model on a test video of VR screen recording and took the average for each model. 
We performed similar evaluation for the Azure API Calls, running each 10 times with test images and averaging the results. \label{test-func}
Upon pressing any of the default interactions (ContextCompass, SceneSweep, and AimAssist), VRSight sends audio feedback to the user (e.g., ``Describing Scene'' in the case of ContextCompass) followed by a description generated by our AI models. This initial audio feedback serves to help mask the response latency to our end-users. We measured the latency of each default interaction using a testing function from initial keypress to the audio playback from our description and report values with and without the audio feedback accounted for.
We calculate the measure with audio feedback included as a ``realized latency'' for the interaction by subtracting the duration of the initial audio feedback from the total response time. 

\subsection{Results}

\begin{table*}[!ht]
\centering
\resizebox{\textwidth}{!}{%
\begin{tabular}{llllllll}
\toprule
Categories & Class Name & Description & Total Instances & \% Total Images & mAP50 & mAP75 & mAP \\ \midrule
\multirow{4}{*}{Avatars} & avatar & humanoid avatar & 7015 & 24.1 & 66.7 & 39.4 & 38.8 \\
 & avatar-nonhuman & non-human avatars, e.g., animals & 1786 & 7.27 & 53.5 & 31.5 & 32.1 \\
 & chat bubble & bubbles adjacent to avatars & 311 & 1.2 & 56.1 & 49.6 & 42.8 \\
 & chat box & chat log showing previous messages & 262 & 1.44 & 89.4 & 71.5 & 65.1 \\ \hline
\multirow{8}{*}{Informational} & sign-text & text-based signs, nametags or portal labels & 14375 & 38.32 & 65.4 & 47.4 & 44.9 \\
 & ui-text & user-only visible text & 7836 & 24.22 & 70.1 & 49.7 & 45.5 \\
 & sign-graphic & world-space signs/posters showing pictures or icons & 6393 & 15.88 & 68.3 & 44.6 & 41.2 \\
 & menu & in-game menu with options. inner buttons unlabeled & 3456 & 16.12 & 78 & 71.4 & 62.2 \\
 & ui-graphic & user-only visible pictures or icons & 2776 & 11.66 & 67.4 & 49.4 & 43.9 \\
 & progress bar & loading screen showing progress & 754 & 4.18 & 53.4 & 28.7 & 31.4 \\
 & hud & “heads-up display”; shows user state e.g., inventory & 394 & 1.9 & 71.3 & 62.9 & 56.3 \\
 & indicator-mute & microphone icon showing mute status & 290 & 1.6 & 77.7 & 68.9 & 57.3 \\ \hline
\multirow{8}{*}{Interactables} & interactable & highlighted or held object, may be grabbable or interactive & 11574 & 28.31 & 63.3 & 45.1 & 41 \\
 & button & pushable object, may trigger interaction & 2572 & 7.81 & 69.9 & 51.8 & 45.9 \\
 & target & object to aim at or hit & 1637 & 3.47 & 78 & 52.9 & 49.7 \\
 & portal & portal for teleporting to virtual worlds & 1182 & 4.48 & 38.3 & 20.8 & 21.1 \\
 & writing utensil & tool for drawing or writing & 788 & 1.76 & 61.8 & 25.1 & 30.5 \\
 & watch & potentially interactive avatar jewelry/watch & 452 & 2.1 & 54 & 32 & 31.7 \\
 & writing surface & surface for drawing or writing & 413 & 1.98 & 68.9 & 59 & 59.5 \\
 & spawner & spawn location for items or avatars & 348 & 1.9 & 62.4 & 52.4 & 49.7 \\ \hline
\multirow{2}{*}{Safety} & guardian & VR boundary - blue/red plus signs; blue grid/cage & 1384 & 5.28 & 67.3 & 58.4 & 56.1 \\
 & out of bounds & growing red circle as player exits boundary & 443 & 1.92 & 86.9 & 71.3 & 69.1 \\ \hline
\multirow{4}{*}{Seating Areas} & seat-single & single-person seating e.g., chairs & 5389 & 10.62 & 69.6 & 52.3 & 48.6 \\
 & table & surface for sitting, placing items or games & 2885 & 13.58 & 70.9 & 62.3 & 55.2 \\
 & seat-multiple & multi-person seating, e.g., couches & 1262 & 5.13 & 57.9 & 38 & 33.5 \\
 & campfire & campfire or fireplace; a likely seating area & 319 & 1.55 & 82.5 & 54.2 & 52.5 \\ \hline
\multirow{4}{*}{VR System} & hand & your hands or other avatars' hands & 12256 & 43.8 & 78.2 & 56.9 & 52.8 \\
 & controller & realistic VR controller model & 3170 & 12.15 & 79.6 & 68.4 & 60.3 \\
 & dashboard & Meta Quest OS menu; bottom bar only & 194 & 1.01 & 84.6 & 59.1 & 56.7 \\
 & locomotion-target & teleportation destination ring on ground & 124 & 0.68 & 28.1 & 10.8 & 14.5 \\ \hline 
Total &  &  & 17691 & 100 & 67.3 & 49.5 & 46.3 \\
\bottomrule
\end{tabular}%
}
\caption{Breakdown of DISCOVR's classes, categorized by the type of VR object. Classes within each category are sorted by number of instances.}
\label{tab:classes}
\Description{A detailed classification table for DISCOVR's dataset of virtual reality objects, organized into six categories: Avatars, Informational, Interactables, Safety, Seating Areas, and VR System. The table provides comprehensive statistics for 30 different VR object classes including their descriptions, total instances, percentage of total images in which they appear, and three detection performance metrics (mAP50, mAP75, and mAP). Notable high-performing classes include "out of bounds" (69.1\% mAP) and "chat box" (65.1\% mAP), while "locomotion-target" shows the lowest performance (14.5\% mAP). The table highlights the distribution and detection effectiveness across different VR interface elements, with the dataset containing 17,691 total labeled instances.}
\end{table*}

\subsubsection{Model Performance}
\label{model-performance}
\begin{figure}
    \centering    
    \includegraphics[width=1\linewidth]{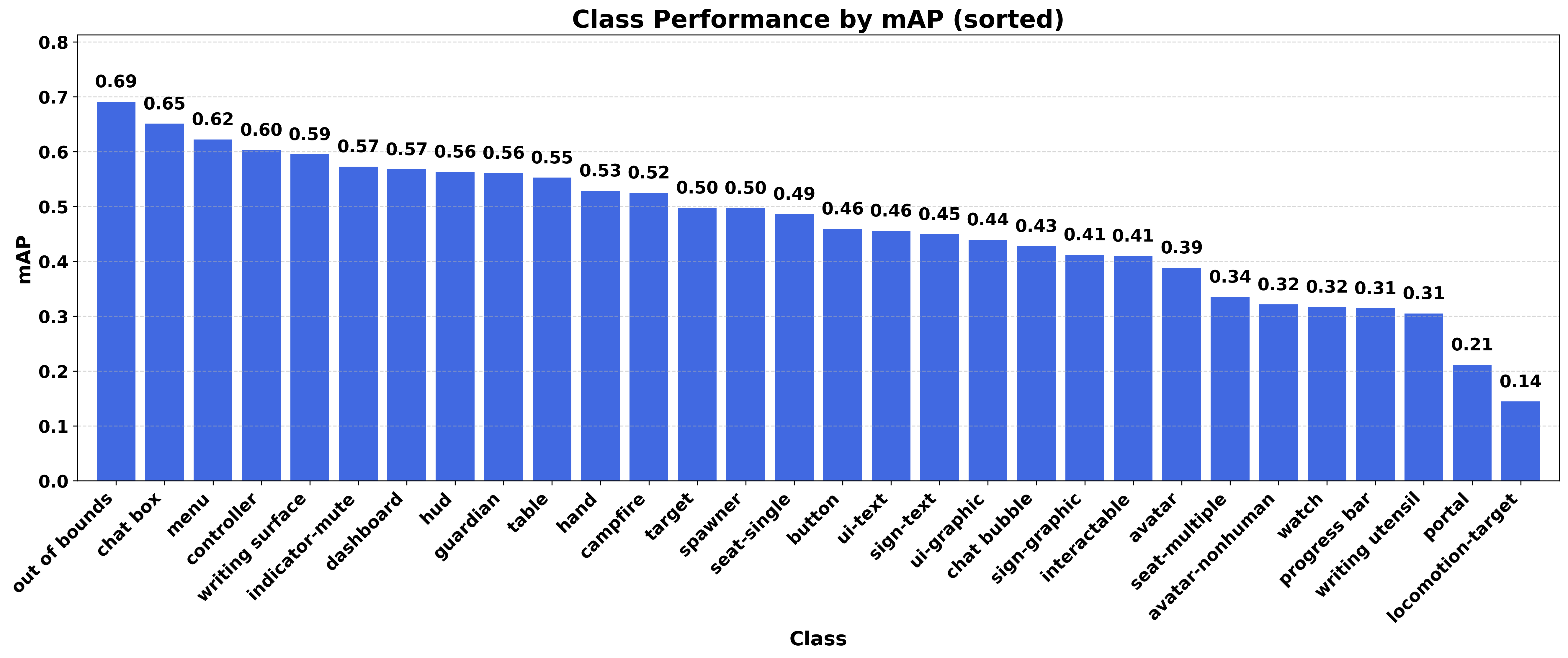}
    \caption{mAP scores (50--95\% IOU) from evaluating each class on DISCOVR's test split. Sorted from highest to lowest.}
    \label{fig:sortedclasses}
    \Description{A bar chart titled ``Class Performance by mAP (sorted)'' showing mean Average Precision (mAP) values for various object detection classes in a computer vision system. The classes are sorted in descending order by their mAP scores, with ``out of bounds'' having the highest score (0.69) and ``locomotion-target'' having the lowest (0.14). The chart displays approximately 30 different classes including interface elements like ``chat box,'' ``menu,'' ``controller,'' and various avatar and interactive elements, with most mAP values ranging between 0.30 and 0.60, indicating varying levels of detection performance across different object categories.}
\end{figure}

Per Ultralytics' validation results, our final model evaluated on the val split at a mAP@50 of 70.6 (average precision that the predicted bounding box and actual bounding box intersect by at least 50\%) and an overall mAP (averaging 50--95 IOU) of 49.7. We further tested the model on the test split, evaluating at an mAP@50 of 67.3, mAP@75 of 49.5, and mAP of 46.3. These values exceed the base YOLOv8n trained on COCO, which reports an mAP of 37.3 for the same 640px input image size~\cite{yolov8_performance_metrics}.
These values are reflected in the Total row of Table~\ref{tab:classes}. Table~\ref{tab:classes} also shows a detailed breakdown of the number of each class present in the dataset, organized by category, and Figure~\ref{fig:sortedclasses} directly compares each class's mAP, sorted from highest to lowest.

\subsubsection{User Input to Feedback Latency}
\textbf{\textit{In Single Query, Optimal Conditions.}}
VRSight operates on a parallel queue system that constantly updates with output from the object, depth, and edge detection models. Should each queue already have a frame pre-processed, using this system end-to-end we measured approximately 2ms of latency: Starting after we register the user's keypress to activating the main function (0.5ms); we then collect the most recent frame from each queue and send an audio packet of the sound data to play at given locations (0.6ms), which our webVR utility (PlayCanvas) then places and plays audibly (1 ms).

\textbf{\textit{Per-Item in Simulated Use.}}
Using the testing function mentioned in \S~\ref{test-func}, we found that our fine-tuned object detection model takes roughly 45.44ms to render each frame (22FPS). Our edge detection is virtually instantaneous, with an average of 0.86ms per frame (1,161.16 FPS). Our depth estimation model DepthAnythingV2-Large is the slowest, rendering a frame every 133.71ms (7.5FPS). 
On querying Microsoft Azure, we receive a response from GPT-4o in 1.69 seconds, OCR in 1.23 seconds, and Text-to-Speech in 0.48 seconds on average.
Each default user interaction calls different models, with ContextCompass querying output from the object detection model, depth model, GPT, and TTS; SceneSweep querying output from the object detection model, depth model, OCR or GPT, and TTS; and AimAssist querying output from the object, depth, and edge models, OCR or GPT, and TTS. 

ContextCompass returns on average 2.501 seconds after keypress. Pressing the ContextCompass key plays a 1.94s audio file reading ``Describing Scene'' for a realized latency of 0.56 seconds. 
SceneSweep returns on average 2.550 seconds after keypress, and its feedback ``Reading all objects'' plays for 2.12s for a realized latency of 0.43 seconds. 
AimAssist returns on average 2.247 seconds after keypress, and its feedback ``Enhanced Object Reading'' plays for 2.20s for a realized latency of 47ms.

Overall, these findings demonstrate VRSight's high accuracy and speed, highlighting impressive detection and playback capabilities.

\section{GENERALIZABILITY ACROSS SOCIAL VR}
\subsection{Procedure}
To further assess the generalizability of VRSight, we evaluated 12 of the VR applications (approx. two-thirds of the original 17 apps identified) following face validity~\cite{nevo1985face}, including VRChat, Engage, Spatial, Zoe, Remio, Meta Horizon Worlds, Rec Room, MeetinVR, ROBLOX, Alcove, Oculus First Contact, and Multiverse. We ran VRSight on each of these VR apps, marking efficacy and challenges we experienced as sighted people.
Each application was evaluated by at least one researcher for at least 20 minutes and at least three different scenes (excluding Oculus First Contact, which only contains one scene). Two researchers evaluated VRChat 
and compared notes to ensure consistency.
Researchers noted VRSight's features in unique scenarios, including unexpected and expected responses from the interaction methods. ContextCompass, SceneSweep, and AimAssist were each used thoroughly for all tests. VRSight was also evaluated on its performance in darker environments, in case the object detection model has poorer performance in poor lighting and darker shading. 
Researchers recorded any issues with the three features in-depth and compared notes to present the following findings. 


\subsection{Results}
\begin{figure}
    \centering
    \includegraphics[width=1\linewidth]{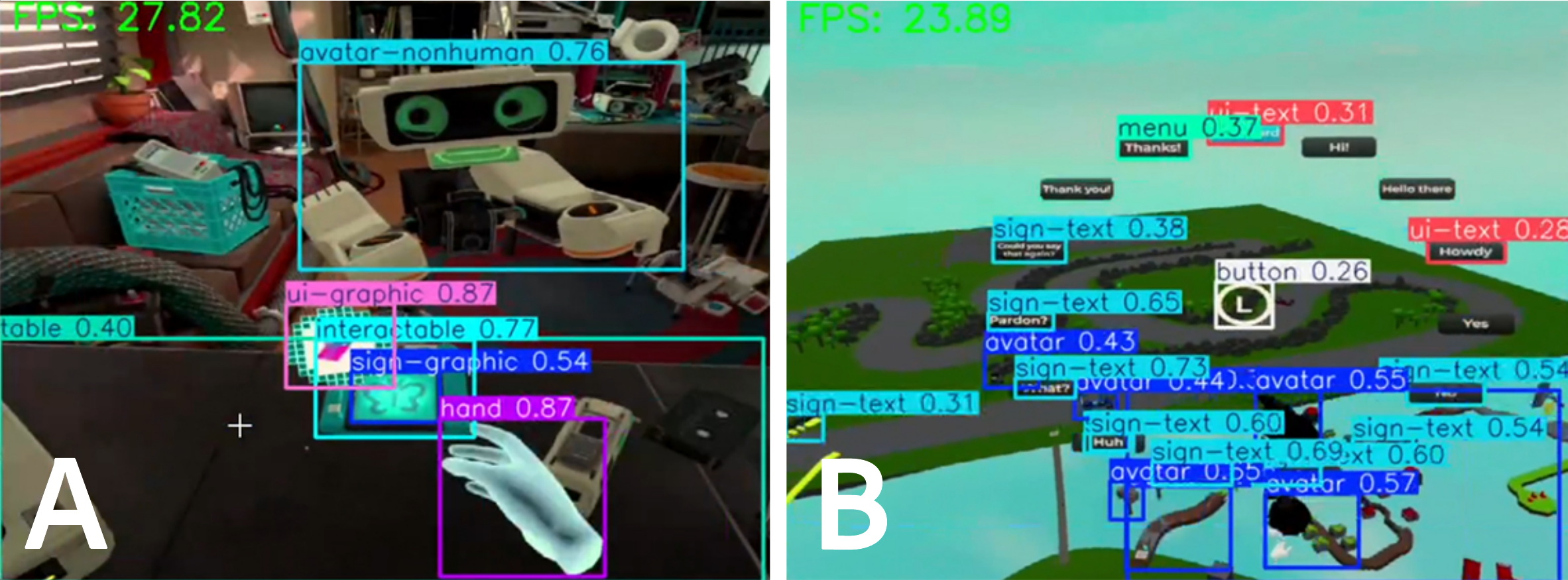}
    \caption{VRSight excels in Oculus First Contact (A), but SceneSweep struggles with many objects on-screen (B).}
    \label{fig:firstcontact}
    \Description{A side-by-side comparison of virtual reality object detection in two different VR environments, labeled as images A and B. Image A shows a physical space with a robot/device (labeled as "avatar-nonhuman") on a table, with additional detected objects including a table, interactable items, sign-graphic, UI-graphic, and a VR hand controller. Image B displays a virtual social environment with multiple avatars and numerous text elements, showing the system's ability to detect and label various interface elements including sign-text labels, UI-text, buttons, menu elements, and multiple avatar representations. Both images include framerate (FPS) information in the upper corners, demonstrating the object detection system running in real-time.}
\end{figure}

\changed{In general, VRSight could be successfully \textit{applied} to all applications in that VRSight was able to detect at least some virtual objects, but performance varied based on factors like reduced detection accuracy in darker VR environments, and lengthy TTS time when many objects are detected.} 
VRSight's object detection model excelled in controlled apps like Oculus First Contact. First Contact's entire app takes place in the same room with the same types of objects, so DISCOVR likely contains nearly complete coverage of the environment (Figure~\ref{fig:firstcontact}.A).

However, we identified some issues that led to the failure of some systems. The object detection model's performance worsened in darker environments (e.g., VRChat, Zoe, Remio) and in lower performant apps (e.g., Multiverse). 
\changed{Triggering SceneSweep when many objects were present functioned but took an extended duration to read all objects (Figure~\ref{fig:firstcontact}.B) or felt repetitive when objects persisted on-screen; for example, MeetinVR contains a Free Trial watermark centered in the user's vision, which SceneSweep perpetually identified and re-read.}

\begin{figure}
    \centering
    \includegraphics[width=\linewidth]{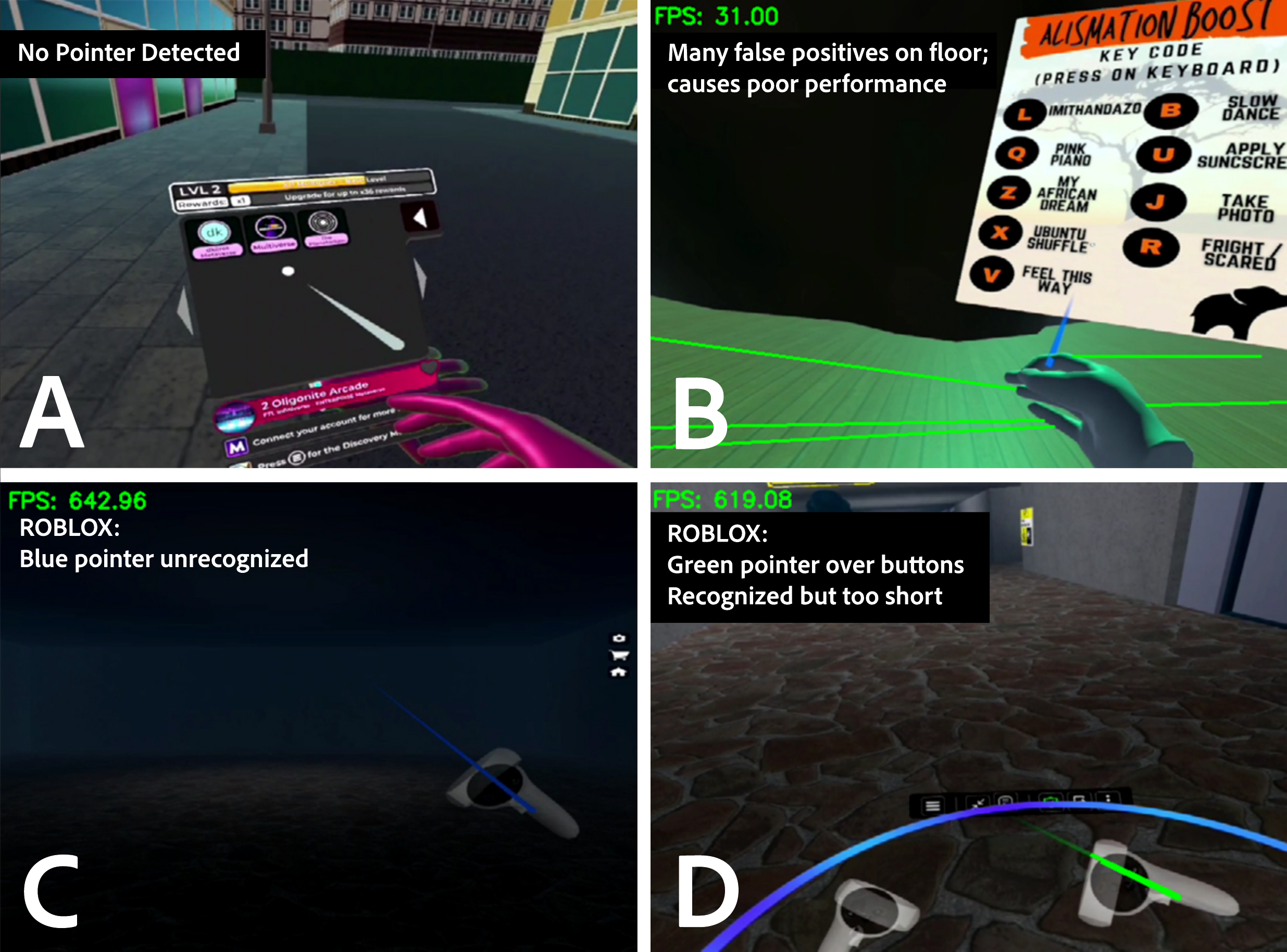}
    \caption{Edge Detection inconsistencies across apps. Image (A) Multiverse's pointer is not detected; Image (B) Spatial's pointer is not detected and there is a great number of false positives in the floor, severely dropping performance; Image (C) demonstrates a dark ROBLOX environment with a blue pointer, not recognized. ROBLOX's pointer turns green when hovering over a valid button, but the detected line does not extend all the way to the appropriate button.}
    \label{fig:edgeexamples}
    \Description{A grid of four virtual reality interface screenshots labeled A, B, C, and D showing pointer detection issues in VR environments. Image A shows "No Pointer Detected" with a ROBLOX interface where a menu system is visible but the pointer isn't recognized. Image B displays "Many false positives on floor; causes poor performance" with an "ALISNATION BOOST" menu board and a green pointer. Image C labeled "ROBLOX: Blue pointer unrecognized" shows a blue ray coming from a controller that the system fails to detect. Image D labeled "ROBLOX: Green pointer over buttons Recognized but too short" shows a VR controller with a green pointer beam that is detected but is insufficient in length to interact effectively with the interface. All images include framerate (FPS) information, with images C and D showing unusually high FPS values of 642.96 and 619.08 respectively.}
\end{figure}

Additionally, as we tuned edge detection for Rec Room's pointers for the user study, the edge detection model did not consistently detect pointers that were not green. Therefore, apps like ROBLOX seldom recognized pointers unless they were already hovering over a valid object, and apps like Spatial and Multiverse always resorted to the fallback behavior; i.e., AimAssist still \textit{functioned} on these apps, but resorted to reading out objects near the user's hand or controller, if detected at all. We provide visual examples for each of these edge detection behaviors in Figure~\ref{fig:edgeexamples}.

\section{USER STUDY}

To supplement our technical evaluations, we conducted an evaluation with nine BLV participants.  
We aimed to evaluate the effectiveness of VRSight to enable people to complete tasks in commercial VR applications and impact on users experience. To this end, we also aim to maximize:
awareness of VR objects (e.g., object detection);
facilitation of user interaction (e.g., edge detection), and 
sense of immersion (e.g., tones, spatial audio).

\subsection{Participants}
We recruited nine BLV participants via our University's mailing list. We screened participants using a demographic questionnaire, requiring that all participants be blind or blind with light perception, or have high visual acuity (VA) of 20/400 or more in their best eye post-correction, and/or limited ($\leq 10$\textdegree) field of view. No existing VR or screen reader experience was required. We recruited six male and three female participants with a mean age of 42.9 ($sd=14.7$). Five participants were blind and four low vision. 
P1 and P8 wore contacts/glasses but were asked to remove them for the study. All participants brought a white cane and/or guide dog except P1 and P7. All participants reported using magnification (P1, P7, P9) or a screen reader (P2--P6, P8) daily. Of the blind participants, four were blind with light perception (P2, P4--P6) while P3 was fully blind. Of the low vision participants, three had limited field of view (P1, P8, P9) and P7 had central vision loss. P7 and P9 also had high uncorrectable visual acuity. Most participants reported little to no VR experience, with two semi-regularly playing VR or spatial audio games for BLV. We enumerate demographic information in Table~\ref{tab:demographics}.

\begin{table}[]
\centering
\resizebox{\linewidth}{!}{%
\begin{tabular}{lllll}
\toprule
ID & Age & Gender & Classification & VR Experience \\ \midrule
P1 & 48 & Male & Low Vision & Plays slow VR puzzle games 1-2x/month \\
P2 & 49 & Male & Blind & None \\
P3 & 25 & Male & Blind & None \\
P4 & 55 & Male & Blind & None \\
P5 & 22 & Female & Blind & Tried VR once but inaccessible \\
P6 & 49 & Male & Blind & Plays BLV audio games 1-2x/month \\
P7 & 49 & Male & Low Vision & None \\
P8 & 26 & Female & Low Vision & Tried VR once but inaccessible \\
P9 & 63 & Female & Low Vision & None \\ \bottomrule
\end{tabular}%
}
\caption{Participant demographic information.}
\label{tab:demographics}
\Description{A demographic table of nine participants (P1-P9) for a study related to visual impairments and virtual reality. The table includes columns for participant ID, age, gender, vision classification, and VR experience. Participants range in age from 22 to 63 years, with six males and three females. Five participants are classified as blind and four as having low vision. Most participants (six) have no prior VR experience, two found VR inaccessible when they tried it once, and two occasionally play VR games (one specifically noting "BLV audio games"). The demographic data suggests this is likely a study examining VR accessibility for people with visual impairments.}
\end{table}

\subsection{Apparatus}
The study was conducted in a well-lit lab environment. Participants sat in an office chair in the center of a 2.5m x 2.5m square space. Next to the space was a desktop computer running VRSight on CUDA with an NVIDIA RTX 4090 GPU. We cast the VR framedata to VRSight using Meta Quest Developer Hub and streamed the frames with OBS Studio set to 640x640 pixel resolution at 90 FPS, downscaling the casting output. We then passed VR visuals over OBS's virtual camera to VRSight. 
Participants wore a Meta Quest 3 VR headset, a pair of standard headphones, and the Quest 3's stock controllers with the three-key keyboard attached to the left controller. The VR headset, headphones, and three-key controller were connected to the computer using long, 4m cables. We collected audio and video recordings using a smartphone and handheld camera, as well as screen recorded each study.

\subsection{Procedure}
The single-session study lasted 2--2.5 hours in-person and consisted of five phases: (1) Demographic questions, (2) a tutorial phase, (3) a task phase, (4) free exploration, and (5) exit questions. Participants were offered breaks when desired and between the task and free exploration section. Participants were instructed to remain seated for the duration of the tasks.
As we focus on post hoc accessibility solutions for off-the-shelf VR apps, no other technologies nor prior work enables similar features we can use as a baseline to compare against.
We hope that others creating VR accessibility features for BLV will find our system useful and use as a point of comparison.
This study was approved by the University of Wisconsin--Madison Institutional Review Board.



\begin{figure}[!ht]
    \centering
    \includegraphics[width=1\linewidth]{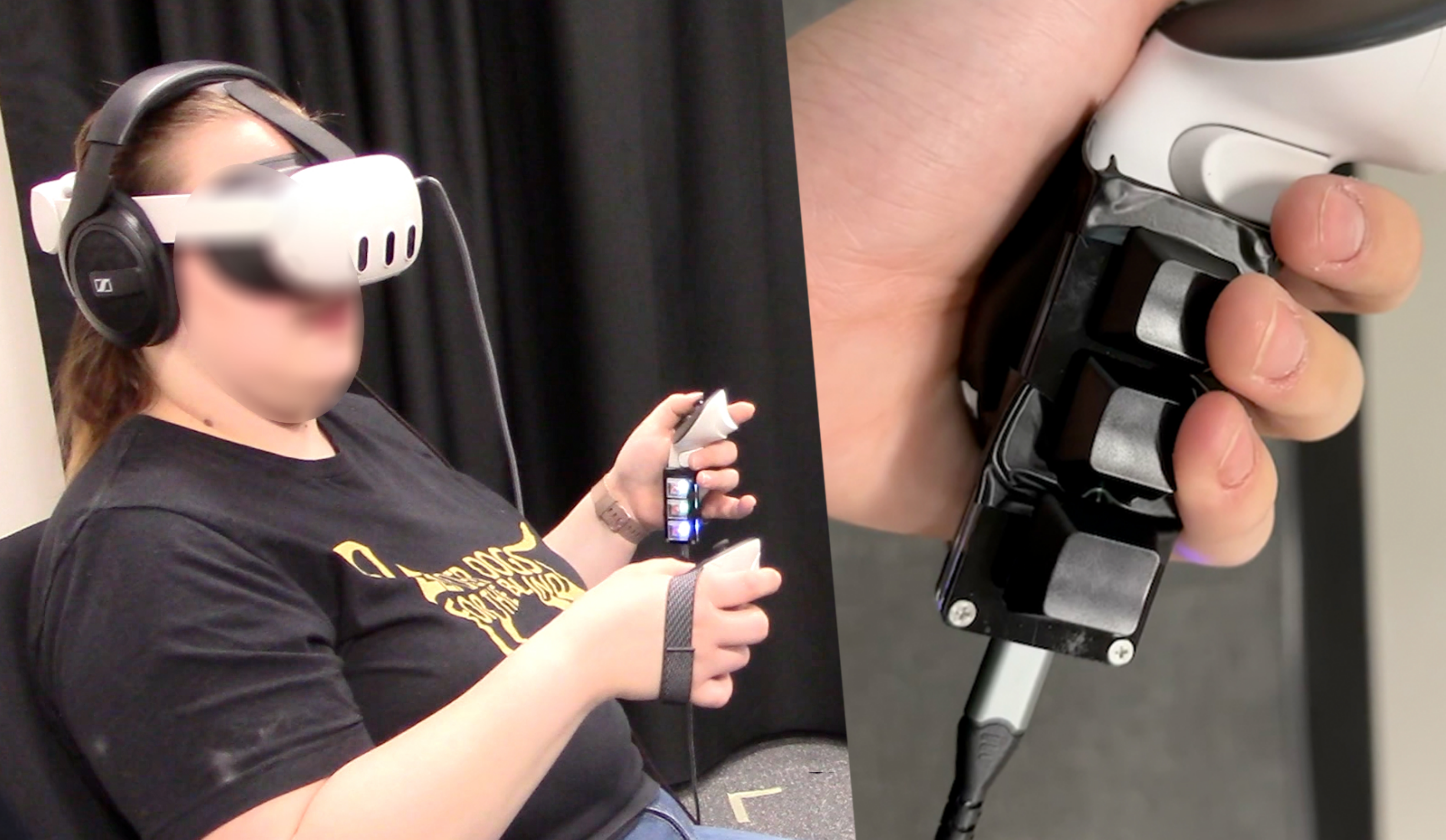}
    \caption{Participant wearing headset (left) using VRSight's three-key controller (right).}
    \label{fig:participantdemo}
    \Description{A side-by-side image showing a study participant using a VR headset and controller. On the left side is a person wearing a white VR headset with over-ear headphones and holding VR controllers, with their face blurred for privacy. They are wearing a black t-shirt with a graphic design and appear to be participating in a research study. On the right side is a close-up view of a hand holding a VR controller, showing the detailed grip and button placement on the device. This image likely documents an accessibility study for VR technology, possibly connected to the research involving participants with visual impairments referenced in the previous table.}
\end{figure}

\subsubsection{Tutorial.} The tutorial tried each of the default interactions using a series of VR test images (e.g., the teaser image). We briefed the participants on each default user interaction, allowing them to experience the type of feedback they will receive from each button press. 
We also helped participants don and adjust the Meta Quest 3 headset and we ensured users could hear the audio playback before proceeding.

\subsubsection{Social VR Tasks.}
\begin{figure}[h]
    \centering
    \includegraphics[width=1\linewidth]{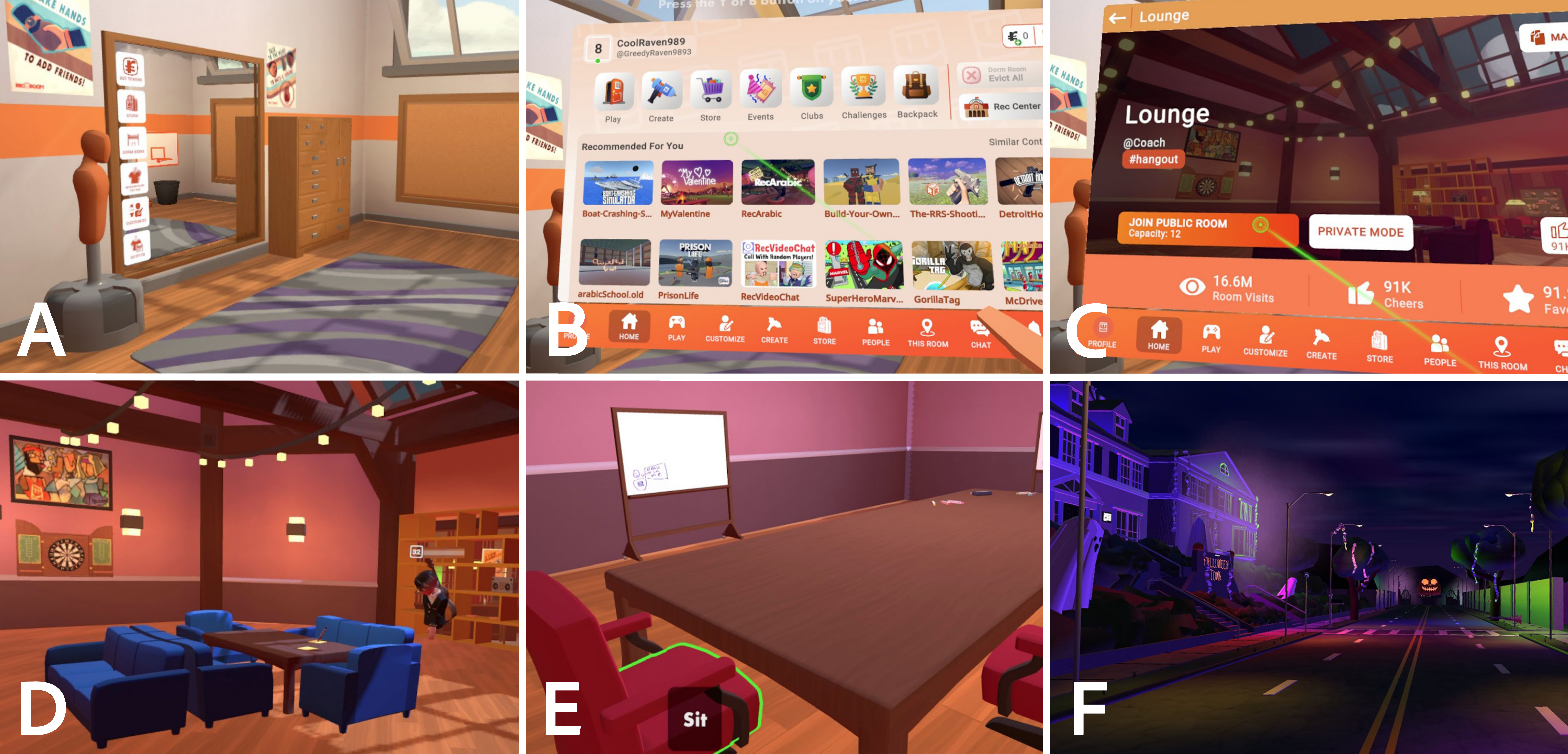}
    \caption{Starting points for each task (A--E) and the free exploration period (F) in order of appearance.}
    \label{fig:starting-points}
    \Description{A grid of six virtual reality/social platform screenshots labeled A through F, showing different environments in what appears to be Rec Room or a similar social VR platform. Image A shows a virtual gym/locker room with an avatar and menu icons. Image B displays a platform home screen with application icons, username "CoolRaven989," and navigation tabs for Play, Create, Store, and other options. Image C shows a "Lounge" space with room statistics (16.6M visits, 91K cheers) and join options. Image D features a cozy virtual lounge area with blue couches and a coffee table. Image E shows a meeting room with a whiteboard and table with a "Sit" prompt visible. Image F displays a nighttime outdoor scene with stylized purple lighting and a Halloween-themed environment. The collection demonstrates various social and functional spaces available in the virtual reality platform.}
\end{figure}

To evaluate VRSight in real-world social VR usage, we devised a set of tasks to explore users' behavior using the popular social VR game Rec Room. All participants explored the same tasks in the same order. For all tasks we disabled locomotion around the space. Incoming audio from Rec Room was enabled, but the participant's outgoing microphone was disabled. All tasks consisted of one trial except the menu task (Task 2) that contained two separate trials. Participants were given a maximum of five minutes per trial and asked to complete each task as quickly and accurately as possible, relying on VRSight when needed. Researchers did not assist participants with any in-scene content. Between each task, we took the participant's controllers, asked post-task questions, and readjusted the player character to match the next task setup. The starting point for each task setup and the free exploration period is enumerated in Figure~\ref{fig:starting-points} in order of appearance. 

\textbf{\textit{Task 1: Explore the Dorm Room (Fig~\ref{fig:starting-points}.A)}}. 
Task 1 positioned participants in the corner near the room's door within Rec Room's starting Dorm Room. 
Participants were asked to identify as many objects in their corresponding directions as possible using VRSight. We used timed each participant's exploration period until they told us they were ready to report their mental map. We evaluated the correctness of their observation by counting the number of correct and incorrect objects identified. We consider an object as ``correct'' if it is listed as the correct description (e.g., poster while pointing in the direction of the poster). We mark the object incorrect if either the object description or direction is wrong.

\textbf{\textit{Task 2: Navigate the Menu (Fig~\ref{fig:starting-points}.B, Fig~\ref{fig:starting-points}.C)}}. 
As VR commonly uses virtual pointers to navigate menus, Task 2 asked users to locate specific objects using their virtual pointer. We broke Task 2 into two trials of up to five minutes each: In the same physical position as Task 1, we opened Rec Room's main menu in the Home tab. The first trial asked participants to navigate to the Backpack Icon and click it with the controller's trigger (Fig~\ref{fig:starting-points}.B). After the first trial, we asked post-task questions, then navigated participants to the page for Lounge, an official Rec Room world. Trial two asked participants to select the orange ``Join Public Room'' button located on the bottom left side of the frame (Fig~\ref{fig:starting-points}.C). 

\textbf{\textit{Task 3: Count the Avatars (Fig~\ref{fig:starting-points}.D)}}. 
In the spirit of a realistic social VR usage scenario, we loaded participants into Rec Room's ``Lounge'' public server. One researcher used a Meta Quest Pro headset to sit in one of the seats within the same room as the participant. Task 3 asked participants to count the number of avatars currently in the room. Since the room was public, some variability occurred with users joining and leaving the room. To analyze responses for this task we standardized the `correct' count by recording the number of avatars in the room upon the participant's last keypress before reporting their score.

\textbf{\textit{Task 4: Find an Open Seat (Fig~\ref{fig:starting-points}.E)}}. 
Our final structured task repositioned participants to the corner of the Lounge's conference table. The researcher in the Quest Pro headset sat in a chair immediately to the right of the participant, and the chair immediately to the left of the participant remained open. Participants were asked to find an open seat and select it using the controller's trigger to sit their avatar down in the open seat. 

\subsubsection{\textbf{Free Exploration (Fig~\ref{fig:starting-points}.F)}}
After completing the tasks we allowed participants to freely explore a Rec Room game called Halloween Town to determine BLV's behavior in a complex social environment. Halloween Town contained a variety of interactive elements, a menu, signs, and dark ambiance. Although VRSight detected the previous tasks as ``neutral'' description tone, VRSight detected Halloween Town as a ``frightful'' scene to ensure varied feedback on the tone detection. We reenabled joystick locomotion using the left control stick and encouraged participants to roll their office chair around the defined playspace, relying on SafeGuard if they navigated too close to the boundary. We asked participants to think aloud their thoughts and recorded common behaviors during this period.

\subsubsection{Igroup Presence Questionnaire.}
\changed{After completing all tasks and the free exploration period, we asked participants to rate statements from the most current iteration of the Igroup Presence Questionnaire~\cite{schubert2001experience,ipq2025} (IPQ) from 1 to 7 to evaluate their perception of the VR environments using VRSight. Rating ``anchors'' for each question varied according to the relevant IPQ question which we assigned to numeric 1 through 7 (e.g., ``fully disagree'' to ``fully agree'' or ``completely real'' to ``not real at all'') with questions INV3 and REAL1 intentionally reversed. We included 13 of the IPQ's 14 questions, removing SP2 as it is purely based on the format of the visual information, not relevant to this study focused on BLV users.}

\subsubsection{Exit Questions.}
We concluded the study with an open-ended interview, asking participants about their experience using VRSight, rating each of VRSight's default interactions\changed{, detection accuracy, and the system as a whole;} giving impressions of the audio feedback; and adding any further improvements they would make to VRSight.

\subsection{Analysis}
We automatically transcribed interviews on-device using iOS Voice Memos. Two researchers independently open coded two transcripts, then met to generate a codebook. The resulting codebook contained 179 codes organized into 15 categories. One researcher then coded the remaining 7 transcripts. The research team met to agree upon the final themes. 

\subsection{Results}
All participants reported benefits of using VRSight, particularly for P1 and P5 comparing the system favorably to their past VR experience. For example P1 reported that ``the description of the chair surrounded by potted plants was really helpful,'' highlighting the system's effectiveness in conveying detailed spatial information.

\textbf{\textit{Task Performance.}}
We evaluated task performance across each structured task and the free exploration period, measuring both interaction frequency and task success. 

\textit{Explore Dorm Room}. 
We classified an object as ``incorrectly'' identified if either the object label was wrong or the direction participants indicated was inaccurate.

In the dorm room exploration task, participants successfully identified numerous objects with high accuracy, demonstrating VRSight's effectiveness for environmental understanding. Participants identified between 4--12 objects ($mean=7.11$, $SD=3.59$) with a low error rate of 17.2\%. ContextCompass proved most useful (mean triggers=5.67, $SD=3.97$), followed by SceneSweep ($mean=4.33$, $SD=2.06$) and AimAssist ($mean=3.89$, $SD=3.33$).
Most participants readily identified key room elements such as posters (\textit{sign-graphic}), chairs, buttons, bunk beds, and windows. The system effectively conveyed spatial relationships, with participants correctly locating objects relative to their position. Six participants completed the task within the time limit (mean completion time=3.15 minutes), demonstrating efficient navigation using VRSight's features. While minor detection errors occurred (e.g., misidentification of doors by P2, P4, P5, P6), these were primarily attributable to either object detection limitations or rarely to language model-generated descriptions that participants trusted (e.g., P8 reporting a ``wooden gaming chair'' that wasn't present).




\textit{Avatar Counting Task.} The avatar counting task revealed VRSight's capability to facilitate social awareness in public virtual environments. Six of nine participants (67\%) correctly identified the exact number of avatars present, with the remaining participants off by 1--2 avatars (mean error=0.67, $SD=0.87$). This high accuracy demonstrates VRSight's potential for supporting social interaction in VR environments. Participants primarily relied on ContextCompass for this task (mean triggers=8.22, $SD=9$), which provided comprehensive scene overviews including avatar presence.
Five participants completed this task within the time limit (mean completion time=3.23 minutes for those who completed), indicating that VRSight provides efficient means for establishing social awareness in VR. The accuracy in this task is particularly notable considering the dynamic nature of avatars, which can move and change appearance, presenting a more challenging detection scenario than static environmental elements.

\textit{Find Open Seat.} The seat-finding task demonstrated VRSight's ability to help users locate interactive objects for physical engagement. Five of nine participants (55.56\%) successfully located and sat in an available chair, showing that VRSight effectively communicated both the presence and functionality of interactive objects. ContextCompass was again the most frequently used feature (mean triggers=5.44, $SD=4.10$), suggesting that general scene descriptions provided valuable contextual information for locating functional objects.
The rapid completion times for successful participants (four completed in under 2 minutes) may highlight VRSight's efficiency for practical navigation tasks. P7 in particular located and sat in the chair in just 22 seconds using VRSight's guidance. P6 found ``blue open seats behind them'', though they had some difficulty with the physical interaction of sitting (pulling the controller trigger at the right location even after hearing the audio feedback for the open seat).



\textbf{\textit{Menu Task.}}
The menu navigation tasks revealed important insights about VR interface accessibility: All participants successfully located the menu in both trials, demonstrating VRSight's effectiveness in identifying UI elements. However, only P7 successfully located the specific backpack icon, and P7 was also the only participant to locate the ``Join Public Room'' button (with P1 and P9 selecting a similar colored but incorrect ``\#hangout'' button).
For menu navigation, participants shifted their strategy, increasing their use of SceneSweep (mean triggers=5.44, SD = 4.45) and AimAssist (mean triggers=12.22, SD = 6.61) in the first trial, suggesting adaptation of their interaction patterns to the task requirements. The challenges participants faced with menu navigation highlight an important area for improvement in VR accessibility, particularly for BLV users who rely on screen readers for interface navigation in other contexts.



\textit{Free Exploration.} During the free exploration period, participants primarily triggered ContextCompass and SceneSweep: ContextCompass (mean triggers=15.67, $SD=10.07$), SceneSweep (mean triggers=11.00, $SD=10.04$), and AimAssist (mean triggers=6.33, $SD=7.28$). 
Eight of nine participants used the full allocated exploration time (10 minutes), with only P6 choosing to end early (3:57). Participants typically began with ContextCompass for orientation, followed by SceneSweep for object identification, and occasionally AimAssist for specific details.

\textit{Across all tasks,} we observed an interesting evolution in feature usage patterns: ContextCompass was consistently the most frequently used feature (mean triggers across all tasks=7.78), indicating its value for general orientation. However, as tasks became more specific (particularly menu navigation), participants increased their use of SceneSweep and AimAssist, suggesting appropriate adaptation of interaction strategies to task requirements.










\subsubsection{Interaction Performance}
We evaluated our three primary interaction methods on scales from 1--7 where seven is very helpful or very easy. Participants largely felt ContextCompass was the most helpful interaction method ($mean=6$, $SD=1$) and easiest to use ($mean=6.78$, $SD=0.44$). SceneSweep was the next helpful ($mean=5.11$, $SD=1.9$) and easy to use ($mean=5.89$, $SD=1.27$) on average. AimAssist received lower ratings for helpfulness but it was acceptably easy to use ($mean=3.44$, $SD=4.33$).
Participants 
rated SafeGuard as quite helpful ($mean=5.8$, $SD=1.1$) and unanimously rated it very easy to use ($mean=7$, $SD=0$). All participants felt the system was helpful, with P9 speaking highly of the feature: ``I wish I had that in real life.'' However, scores varied between 4--7 with some participants unable to easily tell the direction sounds came from, and therefore did not know which direction to move away from---particularly in cases where the guardian was behind or directly in front of them.


\subsubsection{Sense of Presence.}
\begin{figure}[h]
    \centering
    \includegraphics[width=0.8\linewidth]{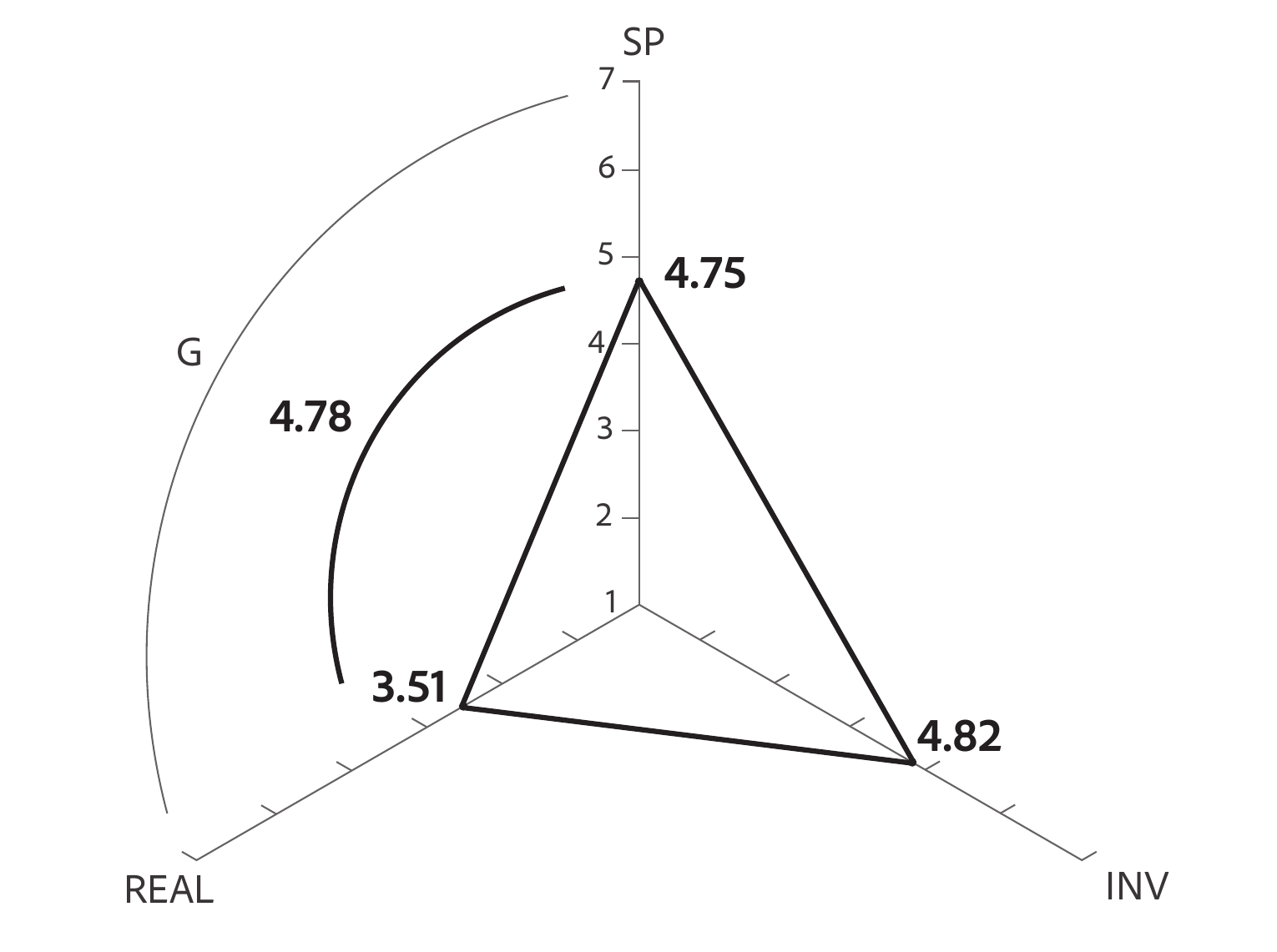}
    \caption{Radar diagram demonstrating participants' average scores from each measure of the presence questionnaire}
    \label{fig:ipq-img}
    \Description{A triangular radar chart/spider plot showing measurements across three dimensions labeled "SP" (top), "INV" (right), and "REAL" (left). The chart displays four numerical values: 4.75 for the SP dimension, 4.82 for the INV dimension, 3.51 for the REAL dimension, and 4.78 for "G" (which appears to be an overall score represented by a curved line). The values are connected to form a triangle, with a scale ranging from 1 to 7 on each axis. This visualization represents assessment metrics for a virtual reality experience, measuring aspects like spatial presence (SP), involvement (INV), and perceived realism (REAL), with G potentially representing a global presence score.}
\end{figure}

\changed{Following the standard presence data analysis for the IPQ~\cite{schubert2001experience,ipq2025} we evaluated the average Spatial Presence (SP1, SP3, SP4, SP5), Involvement (INV1, INV2, INV3, INV4), Experienced Realism
(REAL1, REAL2, REAL3, REAL4), and the general ``sense of being there'' question (G1).}
We found slightly positive results towards VRSight for three of four measures, with Spatial Presence as 4.75 $(SD=1.92)$, Involvement as 4.82 $(SD=1.65)$, and the additional General item as 4.78 $(SD=1.99)$, demonstrating that participants overall felt somewhat present in VR using VRSight. The final measure, Experienced Realism, averaged to 3.51 $(SD=1.95)$ with participants noting that VRSight felt realistic but not necessarily moreso than the real world.
We visualize the average scores in Figure~\ref{fig:ipq-img}, plotting SP, INV, and REAL on three axes and the general item G as a bow to the left.


\subsubsection{Interaction Patterns}
\textbf{\textit{General-to-Specific Approach.}}
Participants often took a top-down approach to self-locate and learn more about their environment; P2 and P9 for example typically started with ContextCompass and worked their way down the interactions to get more specific information about the room around them, using SceneSweep to locate specific objects and occasionally AimAssist for precise targeting. If an interaction gave them an unexpected response, they moved back up to ContextCompass to start over. 

\textbf{\textit{Confidence in General Scene Descriptions.}}
VRSight excelled at providing general scene descriptions, which participants found highly valuable for initial orientation. P5 stated that ContextCompass \textit{``helped me a lot because it described everything to me,''} while P3 noted \textit{``I didn't feel like I was missing anything''} regarding the objects detected. 

Most participants expressed confidence in the system's accuracy. When asked to rate overall accuracy on a scale of \changed{1--7 (very inaccurate to very accurate)}, participants gave an average rating of 5.25 ($sd=1.67$), with P7 and P9 rating accuracy at seven: \textit{``I have to have trust in the system,''} (P7). Most participants could not recall instances where the system misidentified objects, indicating high perceived reliability. 
This effectiveness aligned with our design goal of providing a general-to-specific approach to scene understanding.


\subsubsection{Interface and Feedback Challenges}
\textbf{\textit{Hand-Eye Coordination Difficulties.}}
The virtual pointing system confused several participants. Multiple participants stared at eye level for most of the tasks, for example missing the office chair beneath their eye line (\textit{``I thought it was close...But then couldn't actually locate what it was.''} (P2)) or the edge detection model could not identify their hand during the menu task (P2, P5, P6): \textit{``It's not always obvious to look somewhere to find it. [Turning] your head to explore is not automatically intuitive. [...] I kept forgetting
like, I was wondering what my hands were doing, but I had to look at my hands [to get feedback]''} (P2).

\textbf{\textit{Menu Navigation Challenges.}}
Menu navigation presented similar difficulties, with participants expressing frustration at the lack of dedicated navigation methods outside of precision hand placement, which may have contributed to the low (11\%) menu trial completion rate. 
P7 suggested, \textit{``Just give the person the ability to say read in columns or read in row, start with the left and go across... preserve some of that spatial menu orientation, but still read it out like a screen reader would.''}

\textbf{\textit{Speech Rate and Persistent Feedback Indicators.}}
Participants would like more customizability with speech rate, occasionally experiencing delays between button presses and audio feedback. This behavior created some confusion when multiple controls were activated in sequence, for example when ContextCompass was pressed before SceneSweep finished reading. With SceneSweep in particular, P1 described how \textit{``It started and then it stopped. I'm not sure if I was like waiting for it to keep going because then it went and then it kept going,''} highlighting the need for clearer timing expectations and feedback.

\textbf{\textit{Spatial Audio Enhancement Needs.}}
While participants found spatial audio useful, participants suggested verbal directional and distance cues would enhance orientation. P9 admitted, \textit{``I heard it on the left and the right, but I didn't pay attention to like how far away the object was that they were talking about,''} supporting prior spatial audio research with difficulties in distance.

\textbf{\textit{Mixed Responses to Tone-Based Descriptions.}}
Opinions on tone-based descriptions varied significantly. Most participants found the tone-based analysis interesting, but some participants found the emotional tone distracting. P1 would have preferred a different voice: \textit{``It seemed like it was a little bit extreme... I couldn't really tell between excited and scared.''} Others, like P5, felt \textit{``[The tone] didn't really make a difference... you would have been fine if they sounded the same.''} Several suggested using different voices for different types of information instead of emotional variations of the same voice. P2, P3, and P6 were adamant that the narrator should be as neutral as possible at all times: \textit{``I think of the speech synthesizer is like a neutral agent, so I think when it injects emotion, I think that's distracting''} (P2). Meanwhile, P7 spoke very highly of the tone-based descriptions: \textit{``... People will learn very quickly and are incredibly helpful. There are things that can be done with a narrator voice, tone, pacing, pitch, that communicate things that would take a very long time to describe''} (P7).




\subsubsection{Feature-Specific Findings}
\textbf{\textit{ContextCompass}} received the highest ratings for both helpfulness and ease of use. Participants felt ContextCompass provided reliable, comprehensive scene descriptions. P3 described it as a ``staple button'' to ``refresh and recenter'' himself. P9 appreciated how it ``helped with orienting'' herself in new environments. Several participants suggested a ``specificity slider'' to adjust the level of detail provided between more general to more specific scene descriptions. ContextCompass proved the most helpful, with P1 going so far as to suggest the entire keyboard consist of three levels of specificity of ContextCompass. 

\textbf{\textit{SceneSweep}} SceneSweep was rated moderately helpful and easy to use, with P4 describing it as ``well-designed.'' P9 found SceneSweep ``more helpful than [ContextCompass]''. Some participants suggested improvements in its spatial audio implementation and reliability in different scenes.

\textbf{\textit{AimAssist}} received the lowest helpfulness ratings, with participants finding it difficult to use precisely. Several participants recommended making it toggleable or automatic rather than requiring repeated button presses, tracking the center of the headset position or hand location at all times.

\textbf{\textit{Interface Design.}}
The keyboard interface was acceptable and preferred over alternate input methods like voice or gestures. Participants offered improvements like relocating buttons for easier reach, adding braille to keys (P2), using smaller keys (P2, P7), or using different shaped keys for tactile differentiation (P5, P7). 

\subsubsection{Desired Objects and Features}
When asked about additional types of objects they would like VRSight to identify, participants primarily mentioned objects already in our dataset, including seats, signs/nametags, interactables, avatars, and menus. Several participants (P5, P6, P7, P8) mentioned detection of people or avatars, with P7 suggesting persistent memory about avatar appearances from a friends list. 
Other suggestions included real-world objects not in the dataset (\textit{e.g.,} fire hydrants (P1), mailboxes (P1), furniture (P4), cars/bicycles (P7)), application-specific objects or objectives (\textit{e.g.,} key items (P1), enemies (P4)) or P7's more general suggestion of ``anything that is large and moves''.

Several participants (P4, P7) also expressed interest in bidirectional communication with the AI agent, with P4 suggesting bidirectional communication would enhance the experience significantly, for example specifically asking for the backpack icon and getting ``hot and cold'' as they get closer to or farther away from the target.

\section{DISCUSSION}

Our evaluation of VRSight highlights both its potential for enhancing VR accessibility for BLV  and uncovers several areas for improvement. By combining fine-tuned object detection, tone detection, and spatial audio descriptions, VRSight provides a post hoc solution for making VR more accessible without requiring developer intervention.










\subsection{System Design Implications}
\textbf{\textit{Menu Navigation Design.}}
Our findings suggest that menu navigation requires particular attention in VR a11y systems. P3 equated the VR controller to a Nintendo Wii remote. The Nintendo Wii offered directional-pad style navigation instead of pointer-based input, and participants indicated this would be more effective than pointer-based systems for BLV users. VR app developers should consider incorporating alternate functionality for menu navigation, even remapping the A, B, X, Y buttons to function as a directional pad. Such adaptations would significantly improve menu accessibility and interaction speed for BLV users.

\textbf{\textit{Standardization to improve AimAssist reliability.}}
To improve the reliability of AimAssist, VR app developers should consider standardizing the appearance of VR pointers across applications. VRSight should also include fallback options when pointers are not available or difficult to detect. Interaction systems like NavStick may be applicable if raycasts can be mapped from a top-down perspective to a first-person one in the indicated direction. For objects behind the player, different implementations would likely require source-code access to the VR applications.

\subsubsection{Use of ML to Enhance A11y.}
\textbf{\textit{Multimodal Large Language Models.}}
The object detection system showed promise, particularly in SceneSweep and informing ContextCompass about the types of objects in-frame. Participants found ContextCompass most helpful as it both gave information about objects in and outside the dataset. Using GPT-4o as part of the system enabled us to read contents of icons and use as a fallback to read text when OCR failed. This multimodal approach demonstrates how combining traditional computer vision with state-of-the-art large-language models may enhance a11y for virtual environments with complex visuals.

\textbf{\textit{Multiple Concurrent Object Detection Models.}}
Furthermore, since many social VR worlds incorporate greatly differing themes, art styles, and objectives, VRSight's approach of using a custom dataset specifically for VR objects has proved valuable. However, participants frequently equated the VR environment to their real-world experience, suggesting potential benefit from hybrid approaches. For real-world objects outside of the dataset (e.g., skeuomorphic models), or for applications like Google Earth VR~\cite{google_earth_vr} that blend real and virtual elements, future work may consider running VRSight alongside a supplemental model trained on real-world datasets like COCO.

\subsubsection{System Configuration.} Participants' diverse preferences highlight the need for a11y options to include external configuration menus, including speech rate and tone; description detail level; types of audio feedback; and interaction method preferences. This configurability may address the wide range of user needs within the BLV community to enhance the overall usability of VR and corresponding a11y tools. 

\subsubsection{Enhancing Spatial Audio.} To help BLV better understand virtual spaces, VRSight's spatial audio system should consider methods like Picinali et al.'s acoustic VR platform for the blind to better enable spatial awareness~\cite{picinali2014exploration}, or interaction systems like NavStick to better communicate direction~\cite{nair2021navstick}. 
These approaches could provide richer spatial information beyond basic directional audio. Additionally, incorporating verbal directional cues (``left,'' ``right,'' ``center'') alongside spatial audio would further clarify object positioning, as participants suggested.






\subsection{Limitations \& Future Work}
\subsubsection{\changed{Improving VRSight's Interaction Methods}}
\changed{VRSight comes equipped with a default set of interactions to query the system (\S\ref{interactions}). We invite developers and researchers to devise novel interaction techniques that expand on these defaults for future work on accessible, intuitive interactions for VRSight. Our user study revealed key limitations including menu navigation challenges and customization needs (e.g., speech rate, specificity levels). We believe these limitations could be addressed by incorporating previous works and tools: companion tablet/phone applications for complex menu navigation (TabletinVR~\cite{surale2019tabletinvr}, PhoneinVR~\cite{zhu2024phoneinvr}), adaptive cursors (Bubble Cursor~\cite{delamare2022multifingerbubble,grossman2005bubble}, In-Depth Mouse~\cite{zhou2022depth}) to snap pointer to nearest object, and freezing hand positions to reduce fatigue.}


\subsubsection{On-Device Models and Platform Restrictions}
The current implementation of VRSight requires a separate computer for processing, which introduces some latency and complexity. As computational power in VR headsets continues to advance, we anticipate future systems with sufficient on-device processing capabilities to run these models directly on the headset, reducing latency and simplifying the user experience. \changed{Although VRSight minimally requires a VR headset, standard gaming laptop (verified with an RTX 3070 mobile GPU), and optional three-key keyboard (\$15), achieving our accessibility goal of minimal additional hardware and simple setup, we need an external tool to stream the video feed to VRSight (e.g., OBS Studio). This additional setup step limits BLV users' ability to set up VRSight autonomously, and may require sighted assistance until VRSight is able to connect directly to a background app on-headset.}

Additionally, to play spatial audio on-device, VRSight leverages background audio playback (first introduced in Meta Quest \blue{\href{https://www.meta.com/blog/meta-quest-v66-software-update-reduced-passthrough-distortion-background-audio/}{v66}}) through a webVR utility \textit{(e.g., PlayCanvas)}. 
This method seemingly does not restrict webVR interfaces from continuing an audio stream in the background, but it does prevent accessing the device's IMU sensor data when the playback app is not focused. This behavior makes it difficult to track user head position post hoc, preventing alternate implementations of SceneSweep that may play persistent audio effects as users move their head without attaching additional sensors. We recommend future work investigate alternate implementations that play persistent audio effects as users move their head, potentially requiring modifications to headset firmware or additional external sensors. These hardware restrictions continue to create additional barriers for post hoc accessibility solutions and we urge headset manufacturers to permit their use for accessibility.


\subsubsection{Expanding DISCOVR.}
While our 30 object categories provided good coverage of common VR elements, participants requested additional categories, particularly for app-specific elements or different avatar appearances. \changed{Many participants in our user studies expressed interest in including more objects into DISCOVR, particularly game-specific (e.g., those relevant to the Halloween-themed free exploration area) or real-world outdoor navigation (e.g., fire hydrants (P1)) as participants often received ``No Objects Found'' feedback when exploring with SceneSweep or AimAssist, defaulting back to ContextCompass (GPT rather than DISCOVR's model). Expanding the DISCOVR dataset to include these categories would enhance VRSight's utility in social VR environments. Furthermore, expanding the number of images to DISCOVR above 17,691, increasing the system's confidence thresholds (above the default of 0.25), and/or adjusting hyperparameters to achieve a more precise model may help lower the 17.2\% object identification error rate and increase seat-finding accuracy.
Alternatively, VRSight could use object detection models to improve support for game-specific applications in VR, should they include corresponding models. Given the post hoc nature of the system VRSight \textit{can} run on real-world or mixed reality camera input (e.g., using Passthrough on the Quest 3's cameras), but performance has not been rigorously evaluated.}


\subsubsection{Efficacy Varies Based on App Type and Environment}
\changed{VRSight's post hoc approach as an overlay without requiring developer integration enables broad \textit{compatibility} but introduces application-specific performance constraints. Since VRSight lacks access to underlying game code or scene graphs, its effectiveness varies significantly by application type and environmental conditions.}

\changed{\textbf{\textit{Optimal Performance Contexts.}} VRSight works best in social VR environments with diverse, well-lit interactive objects such as Rec Room and VRChat. These applications provide the stable visual elements and varied object types that our DISCOVR dataset was designed to detect, resulting in reliable object identification and meaningful spatial audio feedback.}

\changed{\textbf{\textit{Performance Limitations.}} Several factors constrain VRSight's effectiveness:
Fast-paced Gaming Environments: VRSight struggles with rapidly changing scenes where objects appear and disappear quickly, leading to stale queries where audio descriptions lag behind the current visual state. This limitation is particularly pronounced in action games or competitive environments.}

\changed{\textbf{\textit{Poor Scene Lighting Conditions:}} Detection accuracy decreases in darker VR environments (e.g., horror games, nighttime scenes) where our computer vision models cannot reliably identify objects due to insufficient visual contrast and detail.}

\changed{\textbf{\textit{Object-Dense Scenes:}} Environments with many simultaneously detectable objects create lengthy text-to-speech sequences that overwhelm users and delay real-time interaction. This issue particularly affects busy scenes or crowded social environments.}

\changed{\textbf{\textit{Application-Specific Elements:}} Our 30-object DISCOVR dataset, while comprehensive for social VR, may miss app-specific UI elements, unique interactive objects, or non-standard visual designs that deviate from common VR conventions or those otherwise unfamiliar to the model.}

\changed{These constraints highlight the trade-offs inherent in post hoc accessibility solutions: while VRSight enables accessibility without developer intervention, it cannot achieve the same level of integration and reliability as native accessibility features built into applications from the ground-up.}

\subsubsection{Variance in AI-Based Description Output}
Although VRSight allowed users to get great descriptions of each scene, self-location within a room remained challenging. Since descriptions were often dynamically generated with AI, repeated keypresses may generate a slightly different description each time. To mitigate against this behavior, systems should include persistent memory or caching behavior to read the same descriptions if participants return to a previously described location, which may both aid participants in mapping the space and increase the system's overall responsiveness.

\subsubsection{Longitudinal Studies with More Participants}
Our evaluation provided valuable insights into initial impressions with VRSight, but longer-term studies \changed{with more participants of diverse backgrounds} would help understand how users adapt to and integrate such systems into VR over time.

\section{CONCLUSION}
In this paper, we present VRSight, a system enabling BLV people to explore visual objects in VR using AI-driven audio descriptions, and DISCOVR, a companion dataset of social VR objects in context. The development of VRSight introduces a revolutionary approach to a longstanding challenge in VR; by integrating advanced object recognition, immersive tone-based audio outputs, and intuitive user interactive capabilities, our system enables BLV to engage with VR environments previously unseen. \changed{We present VRSight as a research demo proving that post hoc AI concepts for VR a11y work; we hope open-sourcing VRSight will encourage further interaction paradigm development using VRSight, alongside motions to combine VRSight's current post hoc implementation with developer-integrated descriptions into the same pipeline.} VRSight helps to bridge the gap between modern technologies and a lack of existing developer support, and we hope VRSight motivates continued VR a11y system development for users and developers alike. 

\begin{acks}
\changed{This work was supported in part by the National Science Foundation under Grant No. IIS-2328182. Thank you to Yuheng Wu for helping select and troubleshoot object detection models, to Abhinav Nandwani for assistance annotating part of the dataset and discussing the initial class list, and to all of our  participants, labmates, and collaborators for their support of VRSight.}
\end{acks}

\bibliographystyle{ACM-Reference-Format}
\bibliography{references}

\appendix


\end{document}